%% file: v2_arxiv_accepted.tex
\documentclass[%
 reprint,
superscriptaddress,
nofootinbib,
 amsmath,amssymb,
 aps,
 prd,
]{revtex4-2}
\usepackage{graphicx}
\usepackage{dcolumn}
\usepackage{bm}
\usepackage{xcolor}
\usepackage{hyperref}
\usepackage{cleveref}
\usepackage{comment}
\usepackage[normalem]{ulem}
\renewcommand{\l}{\left}
\renewcommand{\r}{\right}

\newcommand{\mnras}{Monthly Notices of the Royal Astronomical Society}
\newcommand{\jcap}{Journal of Cosmology and Astroparticle Physics}
\newcommand{\apjl}{The Astrophysical Journal Letters}
\newcommand{\physrep}{Physics Reports}

\begin{document}
\title{Effects of Dark Matter on $f$-mode oscillations of Neutron Stars}

\author{Swarnim Shirke}
 \email{swarnim@iucaa.in (corresponding author)}
\author{Bikram Keshari Pradhan}%
 \email{bikramp@iucaa.in}
 \author{Debarati Chatterjee}
 \email{debarati@iucaa.in}
\affiliation{%
 Inter-University Centre for Astronomy and Astrophysics, Post Bag 4, Ganeshkhind, Pune University Campus, Pune - 411007, India
}%


\author{Laura Sagunski}
\email{sagunski@itp.uni-frankfurt.de}
\author{J\"urgen Schaffner-Bielich}
\email{schaffner@astro.uni-frankfurt.de}
\affiliation{%
 Institut f\"ur Theoretische Physik, Goethe Universit\"at, Max-von-Laue-Stra{\ss}e 1, 60438 Frankfurt am Main, Germany
}%

\date{\today}

\begin{abstract}
The effect of dark matter (DM) on $f$-mode oscillations in DM admixed neutron stars (NSs) is investigated in a comprehensive analysis with particular attention to the role of the nuclear equation of state. 
Hadronic matter is modeled by the relativistic mean field model and the DM model is based on the neutron decay anomaly. 
The non-radial $f$-mode oscillations for such DM admixed NS are studied in a full general relativistic framework. We investigate the impact of DM, DM self-interaction, and DM fraction on the $f$-mode characteristics. We derive relations encoding the effect of DM on $f$-mode parameters. We then perform a systematic study by varying all the model parameters within their known uncertainty range and obtain a universal relation for the DM fraction based on the total mass of the star and DM self-interaction strength. We also perform a correlation study among model parameters, NS observables, in particular, $f$-mode parameters.
Finally, we check the $f$-mode universal relations (URs) for the case of DM admixed NSs and demonstrate the existence of a degeneracy between purely hadronic NSs and DM admixed NSs.
\end{abstract}

\maketitle


\section{Introduction}\label{sec:intro}

NSs are remnants of massive stars that undergo supernova explosion observable throughout the electromagnetic spectrum~\cite{oertel2017, Lattimer2021}. These compact objects are one the densest forms of matter known and observed in the universe. The density inside NSs can reach 2-10 times the nuclear saturation density ($n_0$). They sustain the most extreme physical conditions irreproducible in terrestrial experiments. This, combined with the lack of first principle calculations from the theory of strong interactions, quantum chromodynamics (QCD), makes the interior composition of NSs unknown. NS matter is dense, cold, and highly iso-spin asymmetric. It is conjectured that high densities in the core of NSs can lead to the appearance of new degrees of freedom like hyperons or even result in a phase transition from hadrons to deconfined-quarks~\cite{baym2018, Shirke2023a}. 

In recent years, compact objects have been established as laboratories for studying DM (see~\cite{Baryakhtar2022, Bramante2024} for reviews). DM makes up $\sim25\%$ of our universe and is five times more abundant than ordinary visible matter. DM virializes on galactic scales and interacts with ordinary matter (OM) predominantly via gravity. On smaller scales, DM is known to gravitationally accumulate within condensed bodies like stars and planets~\cite{PressSpergel1985, KraussSrednickiWilczek1986, Gould1987, Gould1988}, although the amount of DM accumulated is only a fraction of the total mass of these objects. NS, being the most compact object after black holes (BH) and hence generating one of the strongest gravitational fields known, is thus expected to be the best candidate for such admixture of DM having larger fractions of DM by mass. A popular mechanism leading to this is the accretion of DM undergoing inelastic collisions with OM within NS, leading to the formation of DM core/halo~\cite{Bramante2024}.The effects of DM core/halo on NS observables have been studied in great detail in the past few years for both Bosonic and Fermionic DM models~\cite{Tolos2015, Rezaei2017, Deliyergiyev2019, ellis2018, Nelson2019, Bhat2020, Das2020, Ivanytskyi2020, DiGiovanni2020, Kain2021, Sen2021, Das2021inspiral, Karkevandi2022, HCDas2022, ADas2022, Dengler2022, Leung2022, Lourenco2022, Miao2022, Collier2022, Dutra2022, Rutherford2023, Hippert2023, Giangrandi2023, Diedrichs2023, Routaray2023, Jockel2024, Mariani2024, Giangrandi2024, Thakur2024, Shakeri2024, Karkevandi2024, Liu2024}. Recently, simulations have been conducted to explore the effect of such DM admixture on the evolution of NSs in binary systems~\cite{Emma2022, Bauswein2023, Hannes2023}. However, such a mechanism cannot lead to substantial DM fractions~\cite{ellis2018}. This is because, DM, if it interacts with other standard model (SM) particles, interacts very feebly and has not been detected so far. The results from DAMA/LIBRA~\cite{bernabei2008DAMA/LIBRAdetection} is the only hint towards a positive detection but is still a matter of debate. Recently, another possibility of neutrons decaying to DM has caught attention~\cite{motta2018a, motta2018b, Husain2022a, Shirke2023b} as it could lead to a large DM fraction~\cite{Shirke2023b, GardnerZakeri2023} in NSs and as well resolve a long-standing discrepancy in particle physics relating to the neutron lifetime~\cite{FornalGrinstein2018prl} called the neutron decay anomaly which is explained below. 

The decay time of neutrons via the $\beta$ decay channel ($n \rightarrow p + e^{-} + \Bar{\nu_e}$) has a discrepancy when measured via two different methods: $1)$ bottle experiments where the number of undecayed neutrons is measured and $2)$ beam experiments where the number of protons produced is measured. The difference in the lifetimes measured in these two methods implies that the number of decayed neutrons is more than the number of produced protons. This problem can be resolved by allowing the decay of neutrons to the dark sector~\cite{FornalGrinstein2018prl}. This model points to new physics beyond the Standard Model and can be linked to the explanations of the dark and baryonic matter asymmetry in the universe~\cite{GardnerZakeri2023}. Applying this idea to NS matter can result in a substantial admixture of DM inside NSs. This makes the neutron decay anomaly model very interesting for NS physics and can have a significant effect on NS observables~\cite{Berryman2022, GardnerZakeri2023}. 
For this reason, we employ the neutron decay anomaly model for DM in the following work. For the hadronic component of NS, we use the well-studied phenomenological relativistic mean field (RMF) model. The microscopic details of these models are described in detail in the next section.

NSs are accessible via electromagnetic observations across the spectrum, right from radio waves to X-rays and gamma rays. EM radiation, originating primarily from the exterior of NSs, provides indirect ways to probe the NS interior. The combination of ground-based and space-based detectors has made numerous measurements~\cite{Lyne2012Book, Ascenzi2024} of NS properties like mass, radius, cooling curves, spin frequency, its derivative, and observed phenomena such as pulsar glitches and mergers, which add several constraints to theoretical models. 
The observed maximum mass of NS imposes stringent constraints on the stiffness of the microscopic equation of state (EoS) that describes NS matter. 
Radius measurements from X-ray observations suffer from model uncertainties and are not precise. The recent NICER mission provides radius estimates to a precision of $5-10\%$ using the pulse profile modeling of X-ray pulses~\cite{riley2019nicermrj0030ads, miller2019nicermrj0030ads, riley2021nicermrj0740ads, miller2021nicermrj0740ads}. Precise simultaneous measurement of mass and radius will highly constrain NS EoS to a high extent. 

Detection of gravitational waves (GW) from the merger of binary neutron stars (BNS), GW170817~\cite{Abbott2017AGW170817} and GW190425~\cite{AbbottGW190425}, and of neutron star-black hole (NS-BH) binaries, GW200105 and
GW200115~\cite{Abbott2021NSBH}, have opened up a new multi-messenger window to study NSs. GW170817 is the first confirmed GW event of a BNS merger that was observed across the electromagnetic spectrum~\cite{Abbott2017AGW170817, abbott2017BGW170817multi, Abbott2017c}. The ability to deduce properties of NSs from GW has renewed interest across a diverse community in astrophysics, as they can also be used to constrain the equation of state and the microscopic properties of NS matter. Precise measurements of NS properties are crucial to determine the interior composition of NSs and the microscopic properties of strongly interacting matter.

On the other extreme, GWs generated due to the time-varying mass quadrupole moment of the entire NS are a direct probe of the NS interior. Analysis of GW170817 added a limit on the tidal deformability ($\Lambda)$ of NSs~\cite{Abbott2019} from the absence of an imprint of the deformation of NSs on the GW signal during the late inspiral phase of the merger, when the tidal field is strong, leading to further constraints on EoS of dense matter~\cite{Abbott2018}. Future observations of NSs from the next-generation GW detector network are expected to improve the constraints significantly. 


In the context of GWs, apart from binary systems, the quasi-normal modes (QNM) of NS are particularly interesting since they carry information about the interior composition and viscous forces that damp these modes. QNMs in neutron stars are categorized by the restoring force that brings the perturbed star back to equilibrium ~\cite{Cowling,Schmidt,Thorne}. Examples include the fundamental $f$-mode, $p$-modes, and $g$-modes (driven by pressure and buoyancy, respectively), as well as $r$-modes (Coriolis force) and pure space-time $w$-modes. The DM admixed NS model that we consider here has been recently studied extensively~\cite{motta2018a, motta2018b, Husain2022a, Shirke2023b}. None of these studies incorporate effects on the QNMs. The effect of admixture of DM on NSs on $r$-mode oscillations was recently studied by some authors in this paper (S. S., D. C., L. S., and J. S. B) for the first time~\cite{Shirke2023b}. It was found that the $r$-mode instability window can be significantly modified if the rate of dark decay is fast enough in dense matter. Several of these modes are expected to be excited during SN explosions, in isolated perturbed NSs, NS glitches, and during the post-merger phase of a binary NS, with the $f$-mode being the primary target of interest~\cite{Kokkotas2001,Stergioulas2011,Pradhan_dyn,Pradhan_ss,Pratten2021,Williams2022,Gamba2022,Vretinaris2020,Ghosh2023}. Among the QNMs of NS, the non-radial $f$-mode strongly couples with the GW emission, and the mode frequency also falls under the detectable frequency range of the current and next-generation GW detectors and holds great importance in NS seismology~\cite{Ho2020,Pradhan_2023apj,Pradhan:2023zmg}. Additionally, there have also been recent works on $f$-mode  GW searches from the LIGO-VIRGO-KAGRA collaboration \cite{AbbottLVK2022,Abbott_2019,AbbottPRD104}. Furthermore, different works have shown that the $g$-modes are less significant than $f$-modes for GW emission ~\cite {Ferrari2003,Lai1999,kruger2015}, leading us to focus on the $f$-mode asteroseismology. 

Recently, some authors of this paper (B.K.P. and D.C.) studied the effect of nuclear parameters and the hyperonic degrees of freedom on the $f$-mode oscillation of NSs in Cowling approximation~\cite{Pradhan2021}, where the perturbations in the background space-time metric are neglected. These results were then improved to include the full general relativistic (GR) effects~\cite{Pradhan2022}. In this work, we extend these studies to $f$-mode oscillations of DM admixed NSs. A recent work~\cite{Das2021} carried out a similar study using a Higgs-interaction model of DM for four select EoS within Cowling approximation. They also highlight the requirement of full-GR treatment for more accurate results, as was also found in~\cite{Pradhan2022}, that Cowling approximation can overestimate the $f$-mode frequencies by up to $30\%$. This was also confirmed by another work~\cite{Flores2024} that appeared during the completion and write-up of the present work. They calculate $f$-mode characteristics in a full-GR setup. However, they consider the Higgs-interaction model and only one fixed nuclear EoS. In this study, we use the DM model based on neutron decay and vary all the model parameters to systematically investigate the effect of DM and its parameters on the $f$-modes oscillations using full-GR. 
Gleason et al.~\cite{GleasonBrownKain2022} dynamically evolved DM admixed NS to study the radial $l=0$ oscillation. However, radial oscillations are known not to emit any GWs and cannot be used to study NS matter. 
In this work, we carry out a systematic study of non-radial $f$-mode oscillations of DM admixed NS in a full GR framework.

This paper is structured as follows: After having outlined the motivation and context of this work in Section~\ref{sec:intro}, we describe the microscopic models for OM and DM along with the formalism to calculate NS observables and $f$-mode characteristics in Section~\ref{sec:formalism}. We present the results of our study in Section~\ref{sec:results} and, finally, summarize our findings in Section~\ref{sec:discussions}.

\section{Formalism}\label{sec:formalism}

We describe the microscopic models used for DM admixed NS matter in Section~\ref{sec:microscopics} and then outline the calculation of their macroscopic properties Sec.~\ref{sec:macroscopics}.

\subsection{Microscopic Models}\label{sec:microscopics}

Here, we describe the particular models we use to describe the hadronic matter (Section~\ref{sec:model_rmf}) and dark matter (Section~\ref{sec:model_dm}) for the study of $f$-modes. We then discuss the choice of model parameters (Section~\ref{sec:params}) we make for the systematic study.

\subsubsection{Model for Hadronic Matter}\label{sec:model_rmf}

The ordinary hadronic matter is described using the phenomenological Relativistic Mean-Field (RMF) model where the strong interaction between the nucleons ($N$), i.e., neutrons ($n$) and protons ($p$), is mediated via exchange of scalar ($\sigma$), vector ($\omega$) and iso-vector ($\boldsymbol{\rho}$) mesons. The corresponding Lagrangian is \cite{hornick2018}
\begin{align}\label{interaction_lagrangian}
\mathcal{L}_{int} &= \sum_{N} \bar\psi_{N}\left[g_{\sigma}\sigma-g_{\omega}\gamma^{\mu}\omega_{\mu}-\frac{g_{\rho}}{2}\gamma^{\mu}\boldsymbol{\tau\cdot\rho}_{\mu}\right]\psi_{N} \nonumber \\
&-\frac{1}{3}bm(g_{\sigma}\sigma)^{3}-\frac{1}{4}c(g_{\sigma}\sigma)^{4} \nonumber \\ 
&+ \Lambda_{\omega}(g^{2}_{\rho}\boldsymbol{\rho^{\mu}\cdot\rho_{\mu}})(g^{2}_{\omega}\omega^{\nu}\omega_{\nu}) + \frac{\zeta}{4!}(g^{2}_{\omega}\omega^{\mu}\omega_{\mu})^{2}~,
\end{align}
where $\psi_N$ is the Dirac spinor for the nucleons, $m$ is the vacuum nucleon mass, $\{\gamma^{i}\}$ are the gamma matrices, $\boldsymbol{\tau}$ are Pauli matrices, and $g_{\sigma}$, $g_{\omega}$, $g_{\rho}$ are meson-nucleon coupling constants.  $b$, $c$, and $\zeta$ are the scalar and vector self-interactions couplings respectively, and $\Lambda_{\omega}$ is the vector-isovector interaction. $\zeta$ is set to zero as it is known to soften the EoS \cite{mueller1996, tolos2017, pradhan2022zeta}. The energy density for this RMF model is given by~\cite{hornick2018}
\begin{small}
\begin{align}\label{energydensity}
\epsilon_{OM}&=\sum_{N}\frac{1}{8\pi^{2}}\left[k_{F_{N}}E^{3}_{F_{N}}+k^{3}_{F_{N}}E_{F_{N}}
-m^{*4}\ln\left(\frac{k_{F_{N}}+E_{F_{N}}}{m^{*}}\right)\right] \nonumber \\
&+\frac{1}{2}m^{2}_{\sigma}\bar\sigma^{2}+ \frac{1}{2}m^{2}_{\omega}\bar\omega^{2}+\frac{1}{2}m^{2}_{\rho}\bar\rho^{2} \nonumber \\ 
&+ \frac{1}{3}bm(g_{\sigma}\bar\sigma)^{3} + \frac{1}{4}c(g_{\sigma}\bar\sigma)^{4} + 3\Lambda_{\omega}(g_{\rho}g_{\omega}\bar\rho\bar\omega)^{2} + \frac{\zeta}{8}(g_{\omega}\bar{\omega})^{4}~,  
\end{align}
\end{small}where $k_{F_{N}}$ is the Fermi momentum, $E_{F_{N}} = \sqrt{k_{F_{N}}^{2} + m^{*2}}$ is the Fermi energy, and $m^{*}=m-g_{\sigma}\sigma$ is the effective mass. Within the mean-field approximation, all the mediator mesonic fields are replaced by the mean values.
The pressure ($P$) is given by the Gibbs-Duhem relation
\begin{equation} \label{eqn:pressure}
P = \sum_{N}{}\mu_{N}n_{N} - \epsilon~,
\end{equation}
where, $\mu_{N} = E_{F_{N}} + g_{\omega}\bar{\omega} + \frac{g_{\rho}}{2}\tau_{3N}\bar{\rho}$. We further have free fermionic contributions from the leptons ($l$), i.e., electrons ($e$) and muons $(\mu)$. This matter is in weak beta equilibrium and charge neutral, resulting in the following conditions,
\begin{equation}\label{eqn:betaeqlbm_neutrality}
    \mu_n = \mu_p + \mu_e,~ \mu_\mu = \mu_e,~ n_p = n_e + n_\mu~.
\end{equation}

For the crust, we use the EoS from Hempel and Schaffner-Bielich (2010)~\cite{Hempel2010} and connect it to the core EoS ensuring causality and thermodynamic consistency.

\subsubsection{Model for Dark Matter}\label{sec:model_dm}

For dark matter, we use a model motivated by the neutron decay anomaly. Fornal \& Grinstein (2018)~\cite{FornalGrinstein2018prl} suggested that the anomaly could be explained if about $1\%$ of the neutrons decayed to dark matter. Multiple decay channels were proposed. Some of these are $n \rightarrow \chi + \phi$, $n \rightarrow \chi + \chi + \chi$, $n \rightarrow \chi + \gamma$~\cite{FornalGrinstein2018prl, Strumia2022}. We consider one of them here, where the neutron decays into a dark fermion with baryon number one and a light dark boson, for which $r$-modes have already been studied~\cite{Shirke2023b}:
\begin{equation}
    n \rightarrow \chi + \phi
\end{equation}
We consider the above decay channel in this work, as it is already well-studied and robust constraints on the self-interaction strength $G$ for this model have been derived in our previous work~\cite{motta2018a, motta2018b, Husain2022a, Shirke2023b, Husain2023b}. The decay channel $n \rightarrow \chi + \gamma$ is known to be ruled out~\cite{tang2018}. For the channel $n \rightarrow \chi + \chi + \chi$, it has been shown that there is no requirement of self-interaction among $\chi$ particles~\cite{Strumia2022, Husain2023a}. The light dark particle $\phi$ with its mass $m_{\phi}$ set to zero escapes the NS, and chemical equilibrium is established via $\mu_N = \mu_{\chi}$. Various stability conditions require the mass of the dark matter particle ($m_{\chi}$) to be in a narrow range of $937.993 < m_{\chi} < 938.783$~\cite{Shirke2023b}. We set $m_{\chi} =  938.0$ MeV. We further add self-interactions between DM particles mediated via vector gauge field $V_{\mu}$.
 The energy density of DM is given by
\begin{equation}
    \epsilon_{DM} = \frac{1}{\pi^2}\int_{0}^{k_{F_{\chi}}} k^2\sqrt{k^2 + m_{\chi}^2}dk + \frac{1}{2}Gn_{\chi}^2~,
    \label{eqn:endens_dm}
\end{equation}
where,
\begin{equation} \label{eqn:Gdefinition}
    G = \left(\frac{g_V}{m_V}\right)^2, \qquad n_{\chi} = \frac{k_{F_{\chi}}^3}{3 \pi^2}~.
\end{equation}
Here, $g_V$ is the coupling strength, and $m_V$ is the mass of the vector boson. From this, we obtain $\mu_{\chi} = \sqrt{k_{F_{\chi}}^2 + m_{\chi}^2} + Gn_{\chi}$. We add this contribution ($\epsilon_{DM}$) to the energy density of hadronic matter ($\epsilon_{OM}$) to get the total energy density ($\epsilon=\epsilon_{OM}+\epsilon_{DM}$) and calculate the pressure using Eq.~(\ref{eqn:pressure}). We vary the baryon density ($n_b = n_p + n_n + n_{\chi}$) and compute the EoS using the conditions in Eqs.~(\ref{eqn:betaeqlbm_neutrality}) and $\mu_N =\mu_{\chi}$.

 our work, we consider neutron dark decay only in the core. The outer crust is mostly composed of nuclei. For the assumed range of the DM particle mass, the decay of neutrons within nuclei is forbidden~\cite{Fornal2020}. Hence, we do not consider any neutron dark decay in the crust. In the inner crust, a fraction of neutrons drip out of nuclei; however, the abundance of free neutrons is much less than that in the core. Therefore, in this work, we neglect the neutron decay in the crust.

We discuss various relevant timescales involving dark decay, chemical equilibrium, and $f$-mode oscillations. The resolution of the neutron decay anomaly model mandates the neutron decay timescale to be 100 times the $\beta$-decay timescale ($\sim 15$ mins), which amounts to a few hours (see~\cite{McKeen2018} for a detailed discussion). This is much smaller compared to the typical age of old NSs $\approx 10^6-10^8$ years. Thus, chemical equilibrium is established in NSs. The typical $f$-mode oscillation frequency for NSs is in kHz, i.e. timescale is in milliseconds. Thus, DM would fall out of chemical equilibrium while undergoing oscillations. For the DM to also undergo oscillations we need to check the timescale of kinetic equilibration between the DM and the ordinary matter via transfer of energy and momentum. Since these fluids interact via gravity, this happens over gravitational timescale which for a typical NS of $M = 1.4 M_{\odot}$ and $R=12$ km is $\sim 1/\sqrt{G\Bar{\rho}_{NS}} = \mathcal{O}(10^{-5})$ s which is less than oscillation timescale. Thus, the system remains in kinetic equilibrium during the oscillations.

Husain et~al.(2022)~\cite{Husain2022a} showed that the decay process could result in a system having temperature of $\mathcal{O}(1)$ MeV which is much smaller compared to the Fermi momenta of either components. Hence, we use $T=0$ EoS for both nuclear matter and DM. During the NS evolution, nuclear matter would cool down to form a degenerate Fermi sea. Since the DM is coupled with neutrons, and establishes chemical equilibrium with it, the DM too eventually forms a degenerate Fermi sea within hours, thus establishing thermal equilibrium well within the NS timescale.

\subsubsection{Choice of parameters}\label{sec:params}
We have a total of eight coupling parameters in this model, six from the hadronic model ($g_{\sigma}$, $g_{\omega}$, $g_{\rho}$, $b$, $c$, $\Lambda_{\omega}$) and two ($g_V$, $m_V$) from the DM model. We set the hadronic couplings using experimental and observational data, as explained below.

The hadronic model couplings are fixed by fitting nuclear empirical data at saturation density. Of these, the iso-scalar couplings ($g_{\sigma}$, $g_{\omega}$, $b$, $c$) are set by the nuclear saturation parameters $n_{sat}$, $E_{sat}$, $K_{sat}$, and $m^*/m$. The iso-vector couplings ($g_{\rho}$ and $\Lambda_{\omega}$) are fixed using the symmetry energy parameters $J$ and $L$. Thus, we fix the nuclear empirical parameters within known uncertainties to generate a particular hadronic EoS. We jointly call the set of nuclear empirical parameters `\{nuc\}'. For the case where we fix the nucleonic EoS and study the variation of $f$-modes with $G$, we fix the nuclear parameters to fixed values as mentioned in Table~\ref{table:parameters}.%
\begin{table*}
    \caption{\label{table:parameters}%
    Range of the variation of the  nuclear and DM parameters used in this work.}
    \begin{ruledtabular}
    \begin{tabular}{c|ccccccc}
       Model& $n_0$ $\rm (fm^{-3})$& $E_{sat}$ (MeV) & $K_{sat}$  (MeV) & $J$ (MeV) & $L$ (MeV) & $m^*/m$ & $G$ (fm$^2$) \\ \hline
        Hadronic & 0.15 & -16.0 & 240 & 31 & 50 & 0.68 & -\\
       Ghosh2022~\cite{ghosh2022multi} & [0.14, 0.17] & [-16.2, -15.8] & [200, 300] & [28, 34] & [40, 70] & [0.55, 0.75] & [0,300]\\
    \end{tabular} 
    \end{ruledtabular}
\end{table*}
We call this case `Hadronic' in this work. The choice of nuclear parameters is made so that the corresponding purely hadronic EoS falls in the chiral effective field theory ($\chi EFT$) band for pure neutron matter, as in \cite{hornick2018}, and forms NS consistent with recent constraints from observational data of maximum NS mass and tidal deformability. This is one set of parameters satisfying these constraints and there is nothing special about it. We choose this as a representative case as the focus is on the effect of DM parameters.  These constraints are described at the end of this section.

Next, to study the correlations and universal relations, we first vary the parameters within the range of uncertainties allowed by nuclear experimental data~\cite{oertel2017, ghosh2022multi, Ghosh2022b} as given in Table~\ref{table:parameters}. We call this range of variation `Ghosh2022' in this work. PREX II experiment suggests higher values of $L$~\cite{Reed2021}. However, we find that such values are inconsistent with the $\chi EFT$ predictions. The same applies to values of $m^*/m$ lower than the given range.

Since the two DM parameters appear as $g_V/m_V$ in the EoS, we explicitly vary only the parameter $G = (g_V/m_V)^2$. In our previous work~\cite{Shirke2023b}, we imposed an updated lower limit on this parameter $G$, demanding consistency with the observation of NSs with a mass larger than 2 $M_{\odot}$. This resulted in a value of $G\gtrsim6$ fm$^2$. This parameter can also be related to the DM self-interaction cross-section($\sigma$), for which we have constraints from astrophysical observations as $0.1 < \sigma/m < 10$ [cm$^2$/g]~\cite{Kaplinghat2016, Tulin2018, Sagunski2021}. This translates to limits on $G$ given by $30 \lesssim G \lesssim 300$ fm$^2$. In the first case, we keep $G>11$ fm$^2$ to keep NS mass larger than 2$M_{\odot}$. For large values of $G$, the DM fraction is observed to be very low, and we do not get any effect of DM. The EoS is asymptotically that of purely hadronic EoS. Thus, in the other case (`Ghosh2022'), where we vary all parameters, we fix the upper limit of the range to 300 fm$^2$. This is also consistent with  $\sigma/m<10$ cm$^2$/g.

To begin, we make some preliminary plots for the model considered. In Fig.~\ref{fig:EoS_diffg}, we plot the EoSs for fixed nuclear parameters and different values of $G$. We use `hadronic' parametrization (see Table~\ref{table:parameters}) for the hadronic matter. The EoS for purely hadronic matter is shown in black. We then add the DM contribution. The EoS is soft when $G$ is low. As we increase the value of $G$, the EoS asymptotically reaches the pure hadronic EoS. This is because the DM fraction decreases with increasing $G$. We show the EoSs with $G=11$, $15$, $30$, $100$, and $300$ fm$^2$.
Self-interaction increases the energy density and makes it energetically more expensive to create DM particles. This is also consistent with the previous study~\cite{Shirke2023b}

  

We only consider those EoSs consistent with $\chi EFT$ at low density ($n_b/n_0 \sim 0.4-1.2$). For any given nuclear parametrization, we generate pure neutron matter (PNM) EoS and check if the binding energy per nucleon falls in the band predicted by $\chi EFT$. If it does, we proceed to generate EoS for the matter with admixed DM.
For every generated EoS, we consider two astrophysical constraints in this work: the corresponding star after solving the TOV equations should have a maximum mass greater than $2 M_{\odot}$~\cite{riley2021} and the tidal deformability (defined in Section~\ref{sec:tov}) of $1.4 M_{\odot}$ star should be compatible with the estimate from the GW170817 event~\cite{Abbott2017AGW170817}, i.e., less than 800~\cite{Abbott2018, Abbott2019} ($\Lambda_{1.4M_{\odot}} < 800$). We call these constraints `Astro' from hereon.

\subsection{Macroscopic Properties  }\label{sec:macroscopics}
In this section, we provide the details of the formalism used to calculate macroscopic NS properties including observables like mass, radius, tidal deformability, DM fraction (Section~\ref{sec:tov}) and $f$-mode characteristics (Section~\ref{sec:fmodes}).

\subsubsection{Calculation of NS Observables}\label{sec:tov}
After varying the parameters and generating EoS, we use this EoS to compute for macroscopic properties of NS like mass ($M$), radius ($R$), and tidal deformability ($\Lambda$). We consider a spherically symmetric non-rotating NS for which the line element is given by $$ds^2 = -e^{-2\Phi(r)}dt^2 +  +e^{2\nu(r)}dr^2 + r^2d\Omega^2~.$$ The macroscopic properties are obtained by solving the Tolman-Oppenheimer-Volkoff (TOV) equations
\begin{align}\label{eqn:TOV}
    \frac{dm}{dr} &= 4\pi r^2 \epsilon(r) \nonumber \\
    \frac{dp}{dr} & = -(p(r)+\epsilon(r))\frac{d\Phi}{dr} \nonumber \\
    \frac{d\Phi}{dr} &= \frac{m(r)+4\pi r^3p(r)}{r(r-2m(r))} 
\end{align}
along with the metric functions~\cite{Tolman1939, OppenheimerVolkoff1939}. In this model, since DM particles are in chemical equilibrium with neutrons, the DM density profiles follow that of hadronic matter, and we get a single fluid-like system. For this reason, we use the single fluid TOV formalism. 

The TOV equations can be solved when supplemented with the EoS $p=p(\epsilon)$. The boundary conditions used while solving TOV equations are $m(r=0)=0$, $P(r=R) = 0$, and $\Phi(r=R) = \frac{1}{2}\log(1-2\frac{M}{R})$. The metric function $\nu(r)$ is given by $e^{2\nu(r)} = \frac{r}{r-2m(r)}$. Thus, by varying the central baryon density, we get different solutions/configurations. $R$ then defines the radius of the stars and $m(r=R) = M$ is the total mass of the star. We do not need to mention a separate central density or DM fraction for DM as the chemical equilibrium with the dark sector fixes the DM density. We calculate the dimensionless tidal deformability $\Lambda=\frac{2}{3}\frac{k_{2}}{C^{5}}$ by solving for the tidal love number $k_2$ simultaneously with TOV equations as done in~\cite{flanagan2008, hinderer2008, DamourNagar2009, YagiYunes2013}. Here, $C$ is the dimensionless compactness $C=M/R$.

The DM fraction is defined as the ratio of the mass of DM in the star to the total mass of the star $f_{DM} = M_{\chi}/M_{tot}$. This quantity is fixed for a given configuration and can be computed as 
\begin{equation}
 f_{DM} = \frac{\int_0^R \epsilon_{DM} dV}{\int_0^R \epsilon dV}
\end{equation}

\begin{figure}
  \includegraphics[width=\linewidth,trim={0 0.5cm 0 0}]{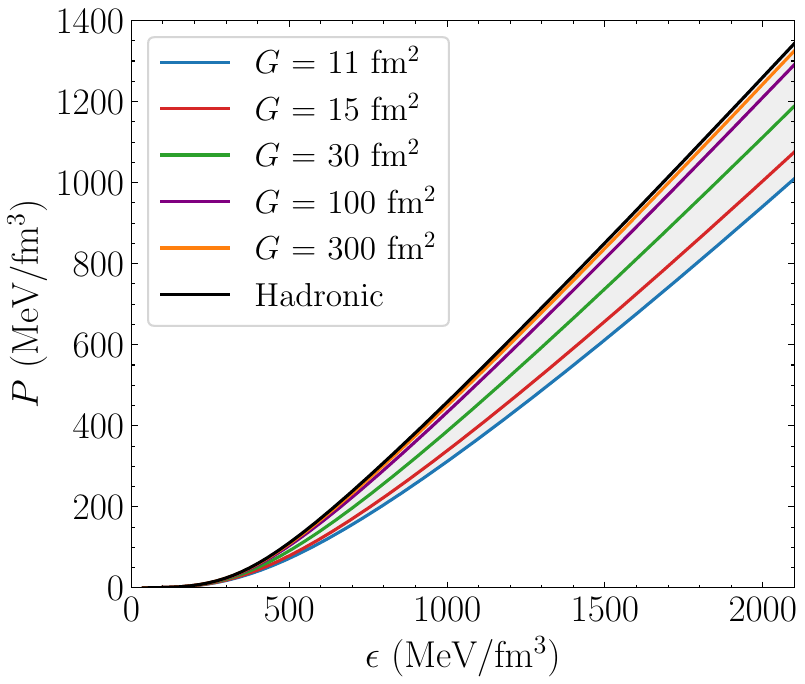}
  
\caption{\label{fig:EoS_diffg}The EoSs for DM admixed hadronic matter with variation of $G$ parameter. `Hadronic' parametrization (see Table~\ref{table:parameters}) is used for the hadronic matter. The black curve denotes the purely hadronic case. }
\label{fig:eos_mr}
\end{figure}

\begin{figure}
  \includegraphics[width=\linewidth,trim={0 0.5cm 0 0}]{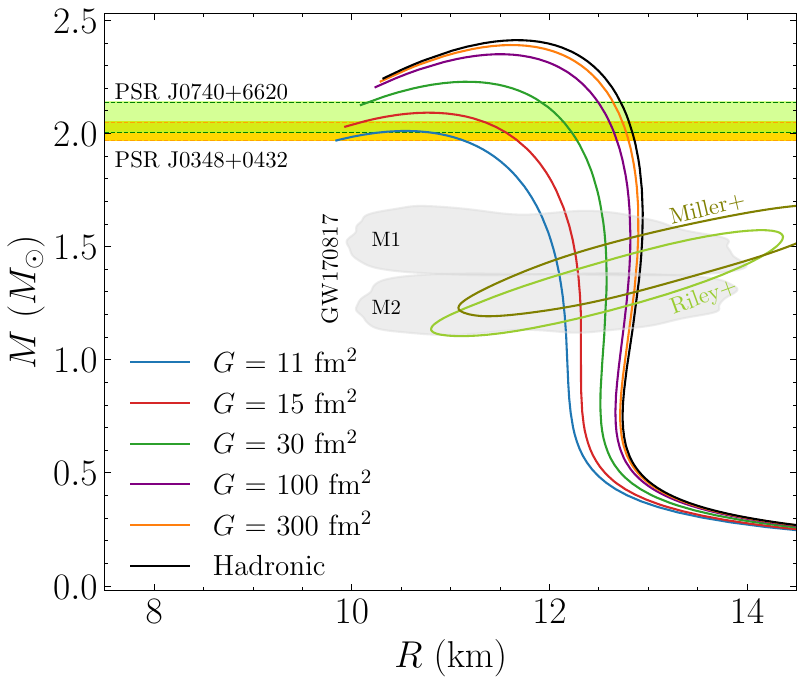}
\caption{\label{fig:MR_diffg}The $M-R$ curves corresponding to EoSs in Fig.~\ref{fig:EoS_diffg}. `Hadronic' parametrization (see Table~\ref{table:parameters}) is used for the hadronic matter. The black curves denote the purely hadronic case. The $1\sigma$ joint $M-R$ contour for the two components (`M1' and `M2') of the GW170817 binary are shown by the gray patch~\cite{Abbott2018}.  `Miller+' and `Riley+' are $1\sigma$ contours derived from the NICER data of PSR J0030+0451 by two independent analyses~\cite{riley2019nicermrj0030ads} and \cite{miller2019nicermrj0030ads}, respectively. The green and yellow bands correspond to mass measurements of the heaviest pulsars known, $M=2.072^{+0.067}_{-0.066}$ of PSR~J0740+6620~\cite{riley2021} and  $M=2.01^{+0.04}_{-0.04}$ of PSR~J0348+0432~\cite{Antoniadis2013}, respectively.}
\end{figure}

For the EoSs shown in Fig.~\ref{fig:EoS_diffg}, the corresponding mass-radius curves is plotted in Fig.~\ref{fig:MR_diffg} after solving the TOV equations~(\ref{eqn:TOV}). The black curve denotes the purely hadronic NS and has a maximum mass of 2.41$M_{\odot}$ and $R_{1.4M_{\odot}} = 12.92$ km. It can be seen that lower values of $G$ lead to configurations with low masses and radii, and the curve approaches the pure hadronic one upon increasing $G$. We show curves for $G>11$ fm$^2$ as they are consistent with the 2-solar-mass constraint. These also agree with the mass-radius constraint from the GW170817 event~\cite{Abbott2018} (gray patch) as well as NICER measurements~\cite{riley2019nicermrj0030ads, miller2019nicermrj0030ads}(green ellipses). We show the bands for the heaviest known pulsars~\cite{riley2021, Antoniadis2013} in the figure for reference. These results are also consistent with our previous study~\cite{Shirke2023b}.

\subsubsection{Calculation of $f$-modes}\label{sec:fmodes}
As indicated by Thorne~\cite{Thorne}, among the various quasi-normal modes of neutron stars (NS), the non-radial fundamental mode ($f$-modes)  serves as a primary source of gravitational wave (GW) emission. Extensive efforts have been dedicated to developing methodologies for determining mode characteristics, including the resonance matching method~\cite{Chandrasekhar:1991}, direct integration method~\cite{Detweiler83,Detweiler85}, method of continued fraction~\cite{Leins1993,Sotani2001}, and the Wentzel–Kramers–Brillouin (WKB) approximation~\cite{Andersson96}. While the relativistic Cowling approximation has been widely used in some studies to find mode frequency by neglecting metric perturbation, several important works~\cite{Yoshida,Chirenti2015,Pradhan2022} underscore the importance of incorporating a linearized general relativistic treatment. These studies conclude that the Cowling approximation overestimates the $f$-mode frequency by approximately 30\% compared to the frequency obtained within the framework of a general relativistic treatment.

In this study, we determine the mode parameters by solving perturbations within the framework of linearized general relativistic treatment. We work in the single fluid formalism and  employ the direct integration method, as outlined in previous works~\cite{Detweiler85,Sotani2001,Pradhan2022}, to solve the $f$-mode frequency of NSs. Essentially, the coupled perturbation equations for perturbed metric and fluid variables are integrated throughout the NS interior, adhering to appropriate boundary conditions~\cite{Sotani2001}. Subsequently, outside the star, the fluid variables are set to zero, and Zerilli's wave equation~\cite{Zerilli} is integrated to far away from the star. A search is then conducted for the complex $f$-mode frequency ($\omega=2\pi f+\frac{i}{\tau}$) corresponding to the outgoing wave solution of Zerilli's equation at infinity. The real part of $\omega$ signifies the $f$-mode angular frequency, while the imaginary part denotes the damping time. Numerical methods developed in our previous work~\cite{Pradhan2022} are employed for extracting the mode characteristics. We refer to Appendix~\ref{sec:appendix_qnm_eqs} for more details of the calculation.

\section{Results}\label{sec:results}
Firstly, we check the effect of the inclusion of DM on $f$-modes. We then study the effect of DM self-interaction on the $f$-mode frequencies and damping timescales, keeping the nuclear parameters fixed. We then vary all the parameters (\{nuc\} + $G$) and check the validity of $f$-mode universal relations for DM admixed NS. Finally, we perform a correlation study to look for any physical correlations.

\subsection{Effect of Dark Matter I: Variation of DM self-interaction}\label{sec:vary_G}
\begin{figure*}
  \includegraphics[width=0.32\linewidth,trim={0.4cm 0.8cm 0.2cm 0}, height=0.27\linewidth]{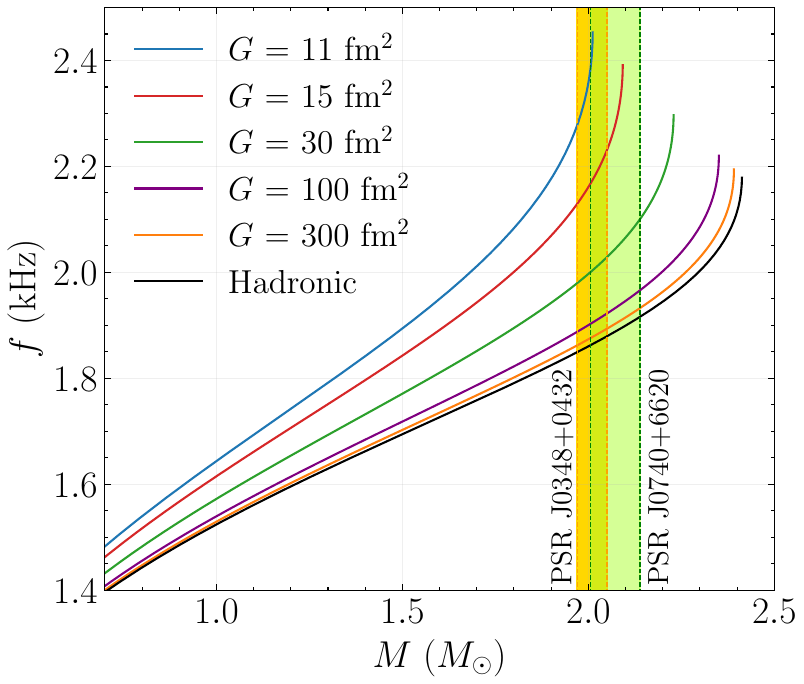}
  \label{fig:f_m_diffg}
  \includegraphics[width=0.32\linewidth,trim={0.4cm 0.8cm 0.3cm 0}, height=0.27\linewidth]{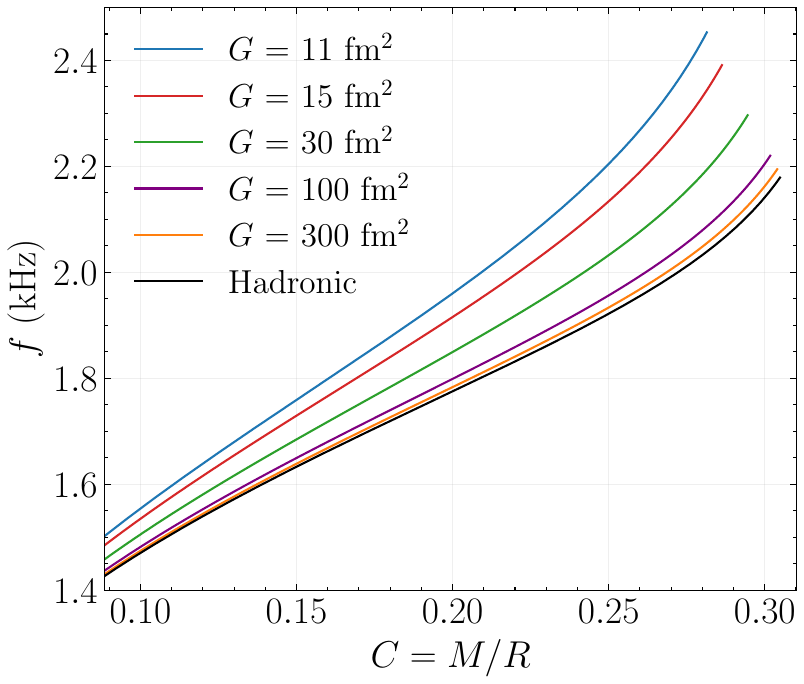}
    \label{fig:f_c_diffg}
  \includegraphics[width=0.32\linewidth,trim={0.3cm 0.8cm 0.3cm 0}, height=0.27\linewidth]{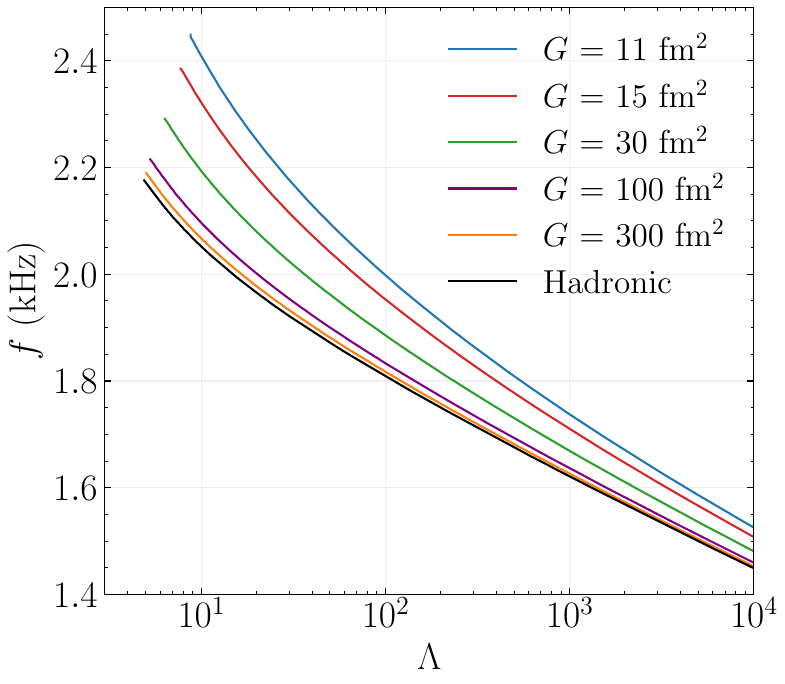}
    \label{fig:f_lam_diffg}
\caption{The $f$-mode frequency ($f$) as a function of (a) NS mass ($M$), (b) compactness ($C$), and (c) tidal deformability ($\Lambda$) for the  EoSs from Fig.~\ref{fig:EoS_diffg} compatible with astrophysical constraints as described in~\ref{sec:params}. `Hadronic' parametrization (see Table~\ref{table:parameters}) is used for the hadronic matter. The black curves denote the purely hadronic case. The green and yellow bands correspond to mass measurements of the heaviest pulsars known, $M=2.072^{+0.067}_{-0.066}$ of PSR J0740+6620~\cite{riley2021} and  $M=2.01^{+0.04}_{-0.04}$ of PSR J0348+0432~\cite{Antoniadis2013} respectively.}
\label{fig:f_diffg}
\end{figure*}

This section focuses solely on the DM self-interaction parameter $G$. To study the impact of the admixture of DM on the $f$-modes, we plot the $f$-mode ($l=2$) frequencies as a function of mass ($M$), compactness ($C$) and dimensional tidal deformability ($\Lambda$) of DM admixed NS in Fig.~\ref{fig:f_diffg}. We use the same EoSs shown in Fig.~\ref{fig:eos_mr}. The bands for the heaviest known pulsars~\cite{riley2021, Antoniadis2013} have been shown in the figure for reference. The black curves represent the purely hadronic case. The maximum $f$-mode frequency corresponding to the maximum mass configuration for the hadronic case is 2.18 kHz, and that for a canonical configuration of 1.4 $M_{\odot}$ is 1.66 kHz. The frequencies increase with mass. We show the $f$-mode frequency profiles for DM admixed NS for selected values of $G$ ($G=11$, $15$, $30$, $100$, and $300$ fm$^2$). 
The inclusion of DM increases the $f$-mode oscillation frequency for a fixed mass configuration. This was also observed in ref.~\cite{Das2021}. The oscillation frequency is higher for denser objects as it scales linearly with the square root of average density (see Section~\ref{sec:UR}). For configurations of fixed total mass, we see that DM admixed NS has a lower radius and, hence, higher average density, leading to higher $f$-mode frequency. We see that as we increase $G$, the frequency reduces. We also observe that the increase in frequency is higher for higher mass configurations.

We see a similar trend when we plot the frequencies against compactness. The frequencies increase with $C$. Compactness is more easily measurable, as the gravitational redshift that the observed thermal X-ray spectrum undergoes depends on the compactness~\cite{Glendenning1997book}. For fixed $C$, we observe that NSs with DM have higher $f$-mode frequencies, which become smaller as we increase $G$. Furthermore, we plot the $f$-modes as a function of $\Lambda$. The frequencies decrease with an increase in $\Lambda$. For a fixed $\Lambda$, the frequency with DM is higher, which decreases with an increase in $G$. Since it is known that the DM fraction ($f_{DM}$) reduces with an increase in $G$, we can conclude that $f$-mode frequency increases with an increase in $f_{DM}$. We will explore this in more detail later in this section.

\begin{figure*}
  \includegraphics[width=0.32\linewidth,trim={0.4cm 0.8cm 0.2cm 0.2cm}, height=0.27\linewidth]{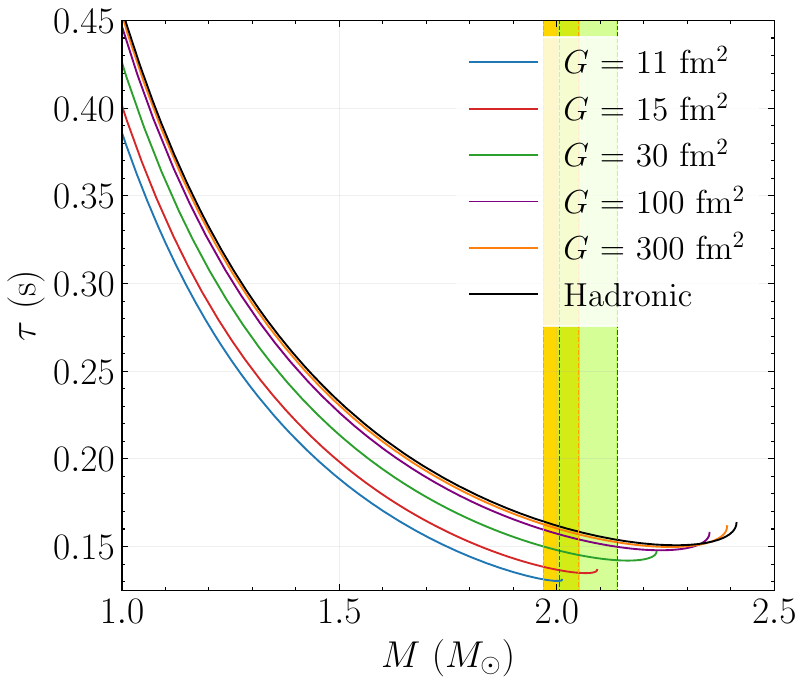}
  \label{fig:tau_m_diffg}
  \includegraphics[width=0.32\linewidth,trim={0.4cm 0.7cm 0.3cm 0}, height=0.275\linewidth]{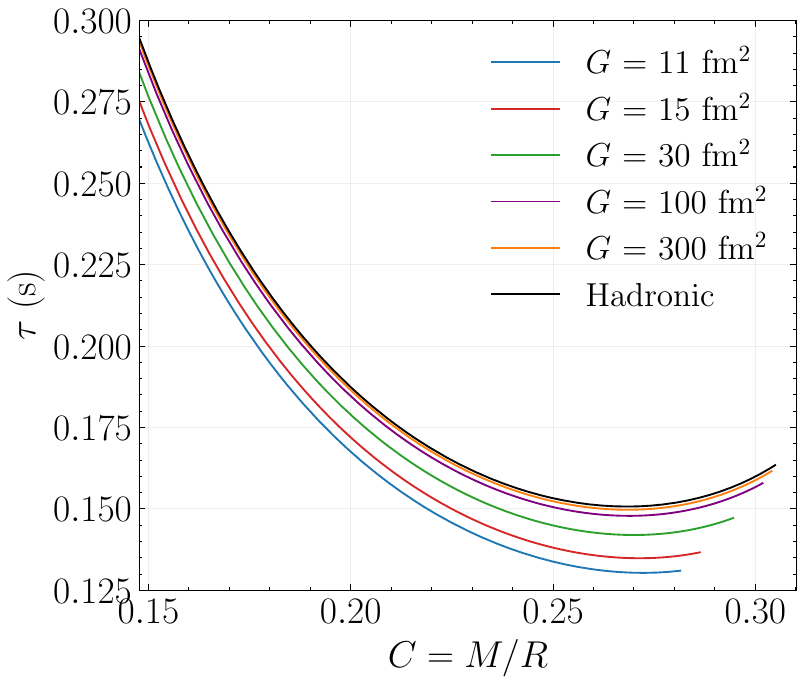}
    \label{fig:tau_c_diffg}
  \includegraphics[width=0.32\linewidth,trim={0.3cm 0.8cm 0.3cm 0}, height=0.27\linewidth]{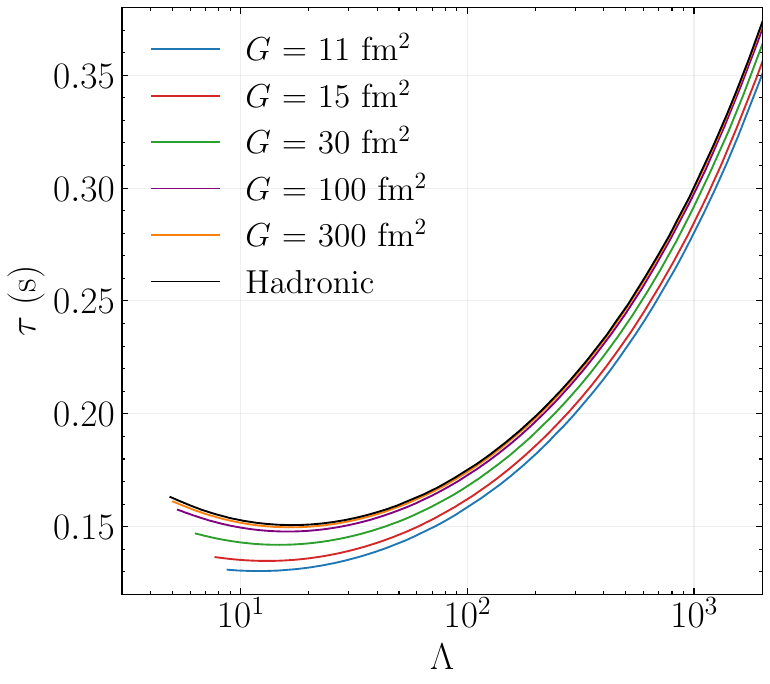}
    \label{fig:tau_lam_diffg}
\caption{The $f$-mode damping time ($\tau$) as a function of (a) NS mass ($M$), (b) compactness ($C$), and (c) tidal deformability ($\Lambda$) for the  EoSs from Fig.~\ref{fig:EoS_diffg} compatible with astrophysical constraints as described in~\ref{sec:params}. `Hadronic' parametrization (see Table~\ref{table:parameters}) is used for the hadronic matter. The black curves denote the purely hadronic case. The green and yellow bands correspond to mass measurements of the heaviest pulsars known, $M=2.072^{+0.067}_{-0.066}$ of PSR J0740+6620~\cite{riley2021} and  $M=2.01^{+0.04}_{-0.04}$ of PSR J0348+0432~\cite{Antoniadis2013} respectively.}
\label{fig:tau_diffg}
\end{figure*}

Parallelly, we also calculate the damping times of these $l=2$ fundamental QNMs for each case. We plot the damping time $\tau$ against $M$, $C$, and $\Lambda$ in Fig.~\ref{fig:tau_diffg}. The black curves denote the pure hadronic case. The $\tau$ corresponding to the maximum mass and canonical $1.4M_{\odot}$ configurations are 0.15 s and 0.26 s, respectively. We see an opposite trend as compared to the frequency. This is expected as $\tau$ is the inverse of the imaginary part of the complex eigenfrequency. The damping time $\tau$ decreases with increasing $M$, $C$ and increases with increasing $\Lambda$. The damping time $\tau$ for DM admixed NS is lower than that of purely hadronic NS. For a configuration of fixed $M$, $C$, and $\Lambda$, the damping time increases with an increase in $G$. We can conclude that the $f$-mode damping time reduces with an increase in $f_{DM}$. $f$-mode frequencies are expected to be detected with good accuracy with the improved sensitivity of GW detectors. However, this is not the case with damping time~\cite{Kokkotas2001}. We explore $f$-mode universal relations in Sec.~\ref{sec:UR}, which can help measure damping time as well.

\begin{figure*}
  \includegraphics[width=0.43\linewidth, height=0.36\linewidth]{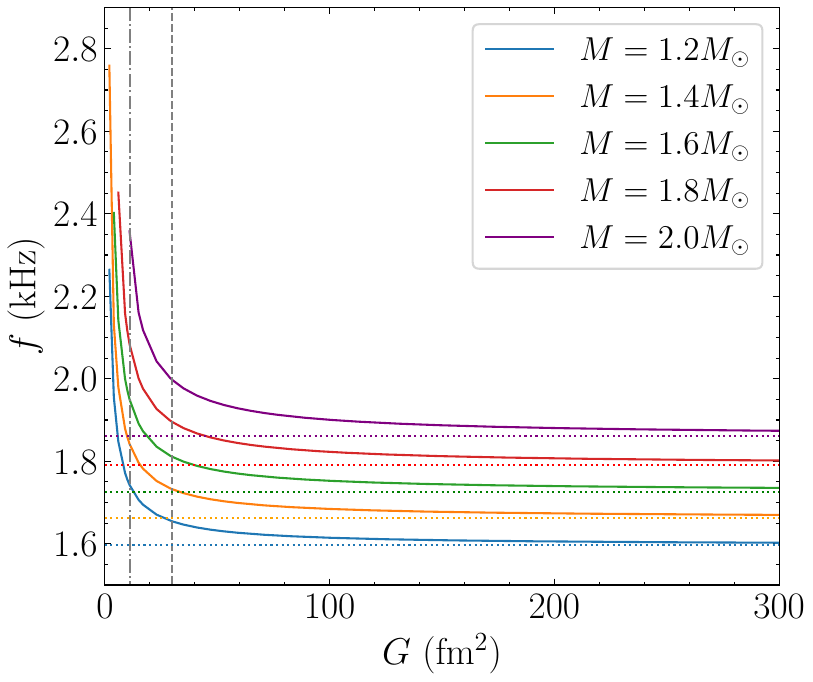}
  \label{fig:f_vsG}
  \includegraphics[width=0.43\linewidth, height=0.36\linewidth]{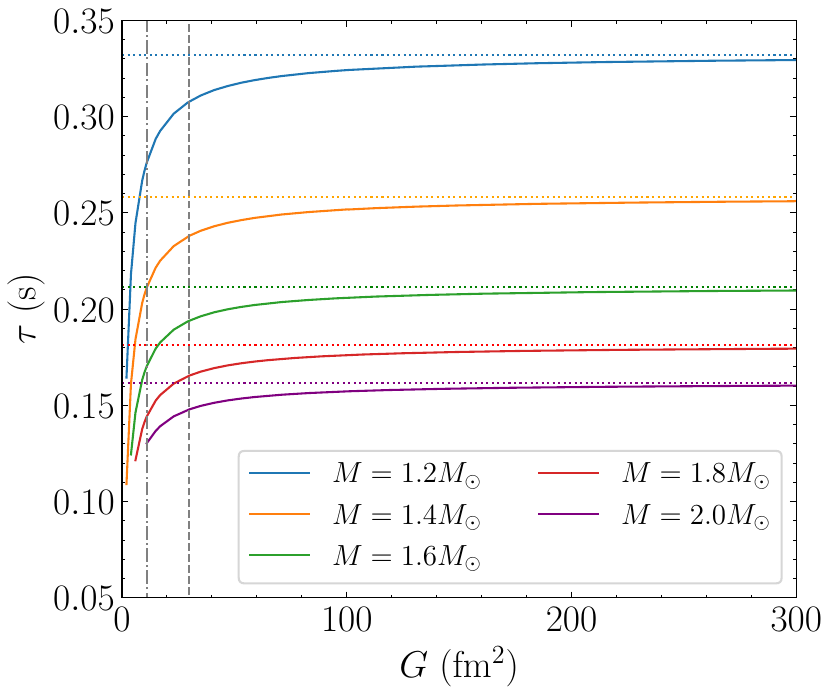}
  \label{fig:tau_vsG}
\caption{(a) $f$-mode frequency ($f$) and (b) damping time ($\tau$) of 1.2, 1.4, 1.6, 1.8 and 2.0 $M_{\odot}$ DM-admixed NS as a function of the self-interaction strength ($G$). The nuclear EoS parameters are as given by the set `Hadronic' in Table.~\ref{table:parameters}. The horizontal dotted lines for each configuration represent the values of the purely hadronic NS. The 2$M_{\odot}$ maximum mass condition for NSs is satisfied by values of $G$ to the right of the vertical dash-dotted line. The vertical dashed line represents the lower limit of $G$ coming from the astrophysical constraint $\sigma/m >0.1$ cm$^2$/g~\cite{Shirke2023b}. }
\label{fig:f_tau_vs_G}
\end{figure*}

We now investigate the effect of DM self-interaction ($G$) in more detail. We keep the nuclear parameters fixed to `Hadronic' (see Table~\ref{table:parameters}) and now vary $G$ continuously. To check the effect of $G$ on $f$-mode characteristics, we plot the $f$-mode frequencies $f_{1.2M_{\odot}}$, $f_{1.4M_{\odot}}$, $f_{1.6M_{\odot}}$, $f_{1.8M_{\odot}}$, and $f_{2.0M_{\odot}}$, for the $1.2M_{\odot}$ (blue), $1.4M_{\odot}$(orange), $1.6M_{\odot}$(green), $1.8M_{\odot}$(red), and $2.0M_{\odot}$(violet) configurations of DM admixed NS as well as their corresponding damping times $\tau_{1.2M_{\odot}}$, $\tau_{1.4M_{\odot}}$, $\tau_{1.6M_{\odot}}$, $\tau_{1.8M_{\odot}}$, and $\tau_{2.0M_{\odot}}$ as a function of $G$ in~\cref{fig:f_tau_vs_G}. The dotted lines indicate the value for the corresponding pure hadronic case for each mass configuration. The vertical dash-dotted line represents the value $G=11$ fm$^2$. Only for $G$ greater than this value do we get configurations that satisfy the 2-solar-mass constraint. The vertical dashed line represents $G=29.8$ fm$^2$ which corresponds to the lower bound on the DM self-interaction cross-section $\sigma/m>0.1$ cm$^2$ coming from astrophysical observations~\cite{Shirke2023b}. We observe that we get high (low) frequencies (damping times) for small values of $G$. The strongest influence of $G$ is in the region $G \lesssim 10$ fm$^2$. The frequency (damping time) falls (rises) upto about $G \sim 50$ fm$^2$ and saturates beyond it. These numbers are not unique as it depends on the hadronic EoS being used. The values here correspond to the `Hadronic' model and are linked to the interaction strength between nucleons via mesons (nucleon-meson coupling). The reason for the sharp (decrease) of $f$-mode frequency (damping time) for $G \lesssim 10$ fm$^2$ is because of the sharp increase in the DM particle fraction in this region. This can be seen from Fig. 7 of~\cite{Shirke2023b}. The DM fraction saturates to zero beyond $ G \gtrsim 50$ fm$^2$ which also explains why the $f$-mode parameters saturate beyond this point.

\begin{figure}
    \centering
    \includegraphics[width=\linewidth,trim={0 0.8cm 0 0}, height=0.7\linewidth]{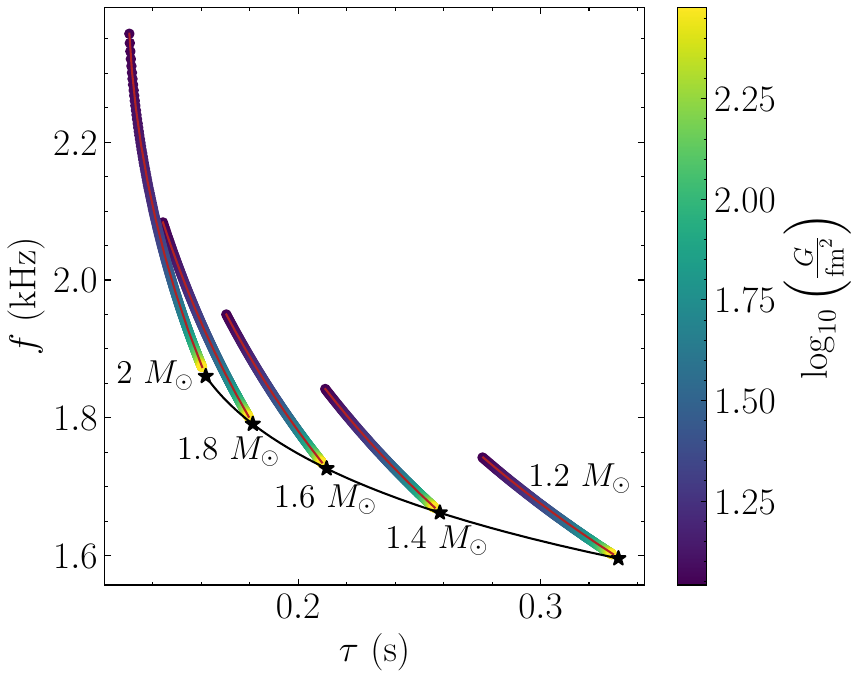}
    \caption{$f$-mode frequency-damping time scatter plot for 1.2, 1.4, 1.6, 1.8 and 2.0 $M_{\odot}$ configurations of DM admixed NS. The colour indicates the self-interaction strength ($G$) on a log scale. The EoS parameters are as given by the set `Hadronic' in Table.~\ref{table:parameters}. The red curves are obtained from the universal relation between $f$ and $\tau$ (See Eq.~(\ref{eqn:UR_f_tau})). The black curve represents the $f-\tau$ curve for purely hadronic EoS (no DM). The black stars on it mark the points corresponding to the mass configurations considered in this plot.}
    \label{fig:f_vs_tau_diffm}
\end{figure}

To see the effect of $G$ on both $f$ and $\tau$ simultaneously for different mass configurations, we make a scatter plot (see Fig.~\ref{fig:f_vs_tau_diffm}) in the $f-\tau$ plane. We show the result for 1.2$M_{\odot}$, 1.4$M_{\odot}$, 1.6$M_{\odot}$, 1.8$M_{\odot}$, and 2$M_{\odot}$ configurations. The colors indicate $\log_{10}{(G/\text{fm}^2)}$ as the variation is resolved better on a log scale. The points for each configuration lie on a curve marked by solid red lines. This is also seen in the case of purely hadronic NSs when the underlying hadronic EoS is varied. 

In our earlier work~\cite{Pradhan2022}, a fitting function was obtained for the mass-scaled $f$-mode frequency and $\tau$, given as 

\begin{equation}\label{eqn:UR_f_tau}
    M\omega_i = \sum_{j} \gamma_j (M\omega_r)^j~,
\end{equation}
where $\omega_r=2\pi f$ is the real part of the eigen-frequency and $\omega_i=1/\tau$ is the imaginary part. Universal relations will be explored in more detail in Section~\ref{sec:UR}. The red curves are plotted using this relation with the fitting coefficients ($\gamma_j$) from \cite{Pradhan2022}, where they were fit for nucleonic and hyperonic matter. We see that the ($f$, $\tau$)-relations obtained when $G$ is varied lie perfectly on the universal relations introducing a degeneracy with nuclear parameters. Thus, simultaneous observation of $f$ and $\tau$ can constrain $G$ only if the underlying nuclear saturation parameters are known to a good precision.

\begin{figure}
    \centering
    \includegraphics[width=0.95\linewidth]{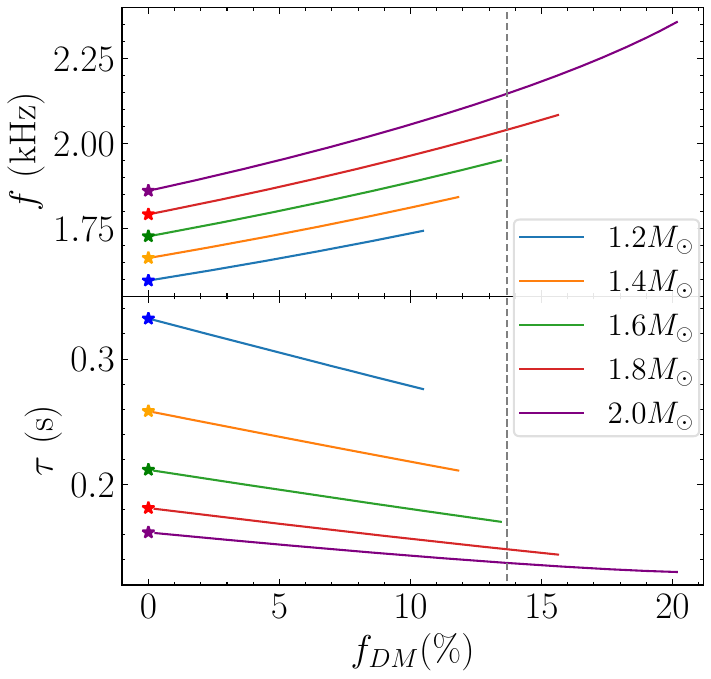}
    \caption{$f$-mode frequency ($f$) and the corresponding damping time ($\tau$) as a function of DM fraction  for 1.2$M_{\odot}$, 1.4$M_{\odot}$, 1.6$M_{\odot}$, 1.8$M_{\odot}$ and 2.0 $M_{\odot}$ configurations of DM admixed NS. The EoS parameters are as given by the set `Hadronic' in~\cref{table:parameters}. The stars represent the value for the purely hadronic case ($f_{DM}=0$). The vertical dashed line represents upper limit on DM fraction ($f_{DM}<13.7\%$) obtained from $\sigma/m >0.1$ cm$^2$/g~\cite{Shirke2023b}.
    }
    \label{fig:ftau_vs_fdm}
\end{figure}

In Fig.~\ref{fig:ftau_vs_fdm}, we plot $f$ and $\tau$ as a function of DM fraction ($f_{DM}$). The configurations shown in this figure correspond to the same curves as in Fig.~\ref{fig:f_tau_vs_G}. The stars shown indicate the purely hadronic case (corresponds to $f_{DM}=0$) for each mass configuration. The vertical dashed line corresponds to $f_{DM} = 13.7 \%$. This is an upper limit of the DM fraction as obtained in our previous work~\cite{Shirke2023b} considering astrophysical constraint $\sigma/m > 0.1$ cm$^2$/g for DM self-interactions. $f$ ($\tau$) is seen to increase (decrease) with $f_{DM}$. This is expected as $f_{DM}$ is known to decrease with increasing $G$. However, in contrast to $G$, we see a linear variation of the $f$-mode parameters with $f_{DM}$.

The lines appear parallel except for a slight deviation for the 2$M_{\odot}$ case for large $f_{DM}$. This can be explained as large $f_{DM}$ corresponds to low value of $G$ and soft EoSs. Since we add a filter of 2$M_{\odot}$, these EoSs have maximum mass near 2$M_{\odot}$. From Fig.~\ref{fig:f_diffg}, it is clear that the variation of $f$-mode characteristics differ near the maximal mass configuration, as mass becomes constant, while $f$ increases. Thus, we expect deviation in trend near 2$M_{\odot}$. The shifts in the lines can be attributed to the difference in $f$-mode frequencies (and damping time) for different mass configurations of the purely hadronic NS. Thus, we define a quantity $\Delta f$ as the difference between the frequency $f$ of a DM admixed NS and that of the purely hadronic NS with the same nuclear parameters given as
\begin{align}
    \Delta f(M, f_{DM}) &= f(M, f_{DM}) - f(M,0) \\ 
    \Delta \tau(M, f_{DM}) &= \tau(M, f_{DM}) - \tau(M,0) 
\end{align}
The dependence of $f$ and $\tau$ on the underlying microscopic parameters (\{nuc\}) and $G$ is implicit in these equations. The dependence on $G$ is only through $f_{DM}$. $f_{DM}$ also depends on $M$. We will explore these relations in detail later. Also, at this stage, we cannot say whether $\Delta f$ and $\Delta \tau$ depend on the nuclear saturation parameters. 

\begin{figure}
    \centering
    \includegraphics[width=0.95\linewidth,trim={0 0.5cm 0 0}, height=0.9\linewidth]{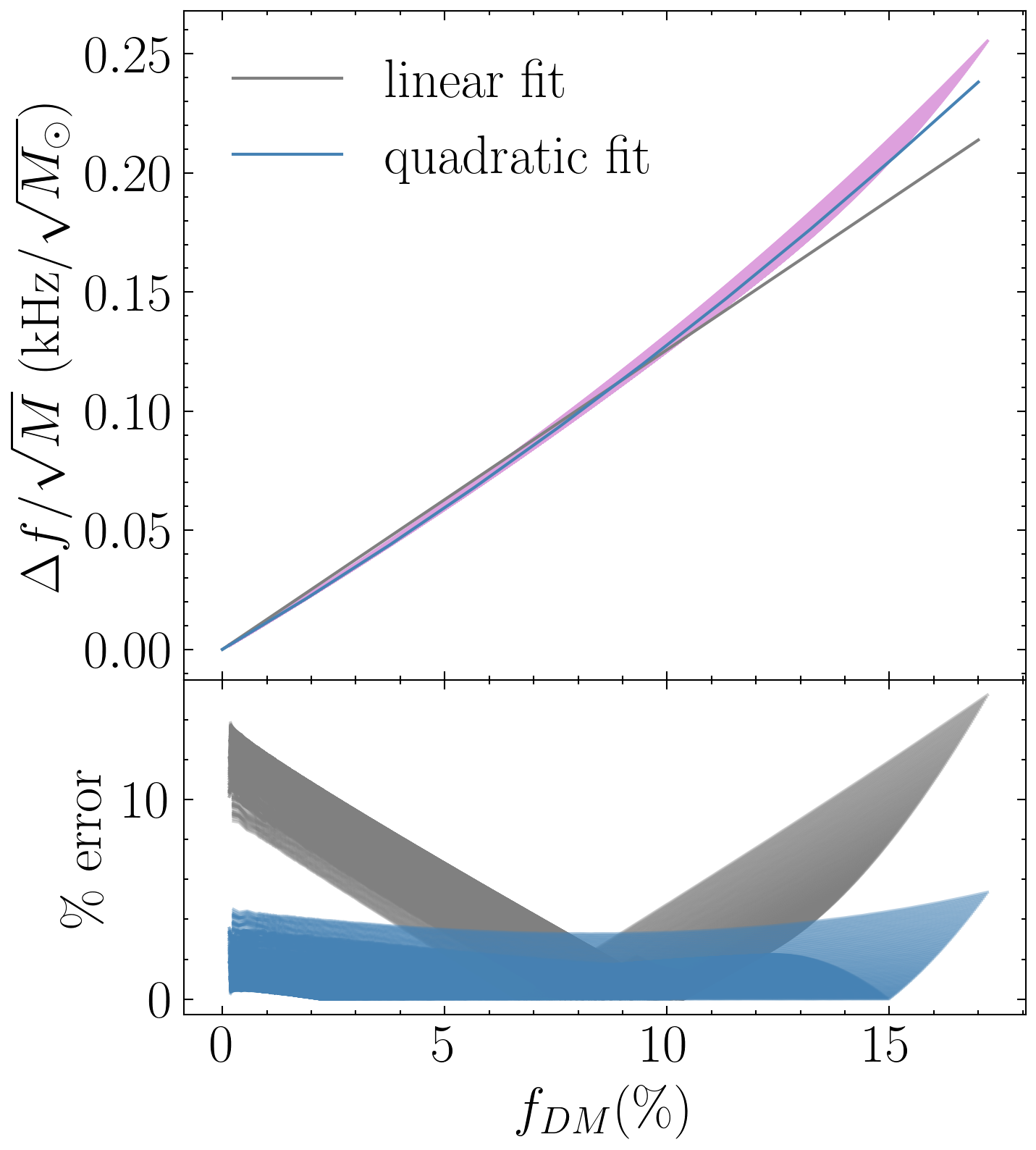}
    \caption{The top panel shows $\Delta f(M_i)/\sqrt{M_i}$ as a function of $f_{DM}$ obtained by varying $G$ and fixing the nuclear parameters to `Hadronic' (refer Table.~\ref{table:parameters}). The gray and blue lines are linear and quadratic fits given by Eqs.~(\ref{eqn:deltaf_fit_linear}) and (\ref{eqn:deltaf_fit_quadratic}). The fit coefficients are reported in Table.~\ref{table:deltaf_deltatau_fits}. The bottom panel shows percent error for the two fits.}
    \label{fig:deltaf_vs_fdm_fit}
\end{figure}

\begin{figure}
    \centering
    \includegraphics[width=0.95\linewidth,trim={0 0.5cm 0 0}, height=0.9\linewidth]{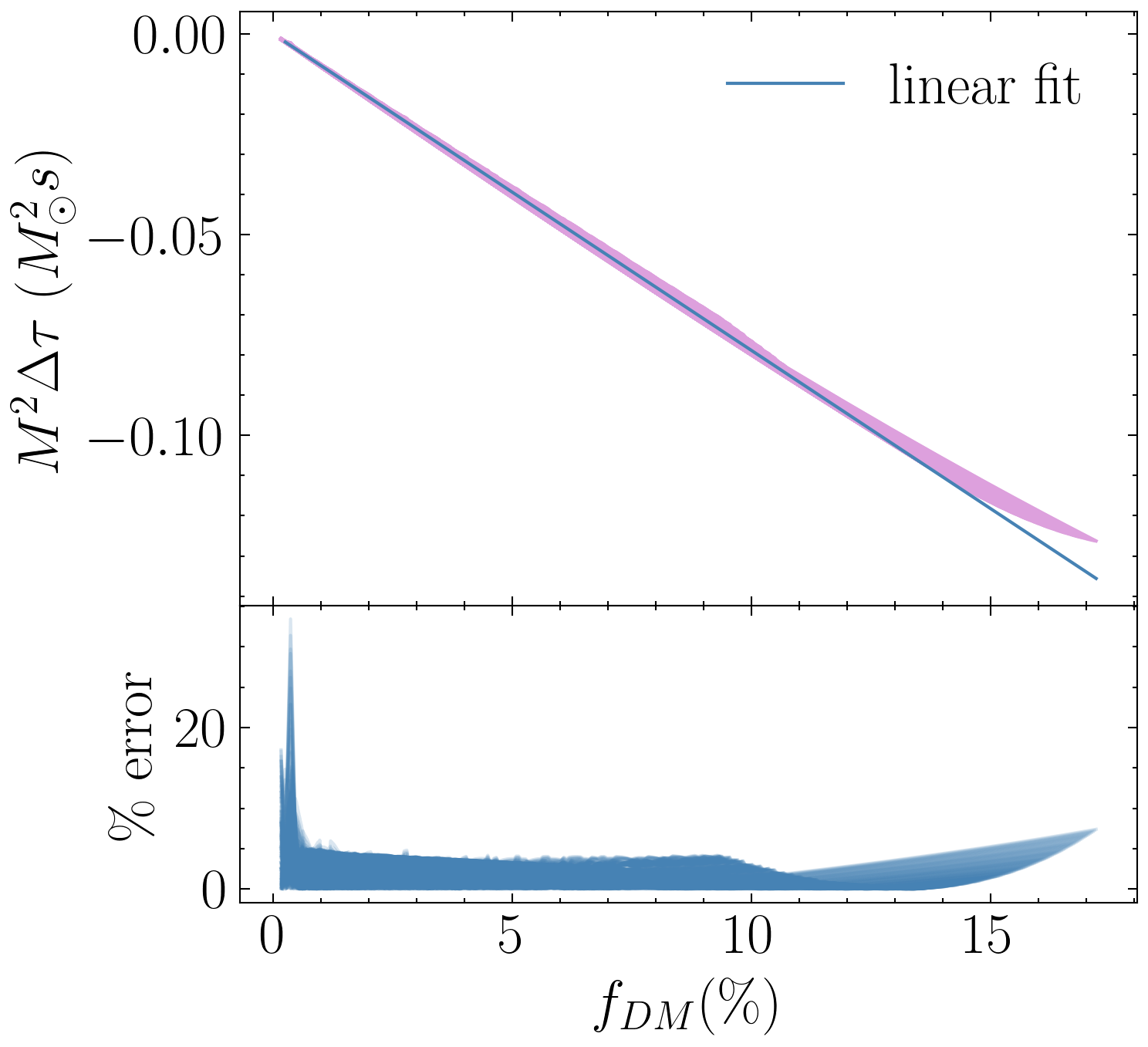}
    \caption{The top panel shows $M^2\Delta \tau(M)$ as a function of $f_{DM}$ obtained by varying $G$ and fixing the nuclear parameters to `Hadronic' (refer Table.~\ref{table:parameters}). The blue line is a linear fit given by Eq.~(\ref{eqn:deltatau_fit}). The fit coefficients are reported in Table.~(\ref{table:deltaf_deltatau_fits}). The bottom panel shows the percent error for the linear fit.}
    \label{fig:deltatau_vs_fdm_fit}
\end{figure}

When we plot $\Delta f$ and $\Delta \tau$ as a function of $f_{DM}$ (not shown here), we obtained straight lines with different slopes for different mass configurations. Analyzing the effect of mass, we find that the slope is proportional to $\sqrt{M}$ for $\Delta f$ and $M^{-2}$ for $\Delta \tau$. Thus, we expect $\Delta f/\sqrt{M}$ and $M^2\Delta \tau$ to fall on a  straight line. To test this, we take about 50 EoSs corresponding to different values of $G$  uniformly spaced between 11 fm$^2$ and 300 fm$^2$. All these EoS are consistent with the constraints considered in this work. As we discussed, there is a deviation of trend near 2$M_{\odot}$, so we restrict to the mass range of [1,1.9] $M_{\odot}$ while studying these relations. We take 500 mass values ($M_i$) within this range and compute $\Delta f(M_i)/\sqrt{M_i}$ and $M_i^2\Delta \tau(M_i)$  as a function of $f_{DM}(G)$.  We plot these in Fig.~\ref{fig:deltaf_vs_fdm_fit} and Fig.~\ref{fig:deltatau_vs_fdm_fit} respectively. 

Fig.~\ref{fig:deltaf_vs_fdm_fit} shows that we get a tight relation between $\Delta f/\sqrt{M}$  and $f_{DM}$. %
We perform linear and quadratic fits to it, given by
\begin{align}
    \Delta f(M, f_{DM}) &= \sqrt{M}(C_{f1} f_{DM}[\%]) ]~, \label{eqn:deltaf_fit_linear} \\
\Delta f(M, f_{DM}) &= \sqrt{M} (C_{f2}  f_{DM} [\%] + C_{f3} (f_{DM} [\%])^2)~. \label{eqn:deltaf_fit_quadratic}
\end{align}
$C_{fi}$ are the fitting parameters. $M$ is in units of $M_{\odot}$, and $f_{DM}[\%]$ is the percentage of DM fraction. We impose the condition that for $\Delta f(M, f_{DM}=0)= 0$, i.e., $f_{DM}$ should correspond to purely hadronic NS. This fixes the zeroth order term, independent of $f_{DM}$, to zero, which then has not been considered in the fit. The fit coefficients are reported in Table.~\ref{table:deltaf_deltatau_fits} along with the coefficient of determination ($R^2$). The bottom panel shows the absolute percent error (defined as $100\times|\Delta q/q|$ for any quantity $q$). We see the linear curve fits to an accuracy of $15\%$. This relation can be used to estimate the increase in f-mode frequency of a DM admixed NS for any given mass configuration and DM fraction. Using the fitting coefficient, we can approximate $\Delta f(M, f_{DM}) \approx 1.3\sqrt{M}f_{DM}$. We improve the fit by considering a quadratic function and get a tighter relation with an accuracy within $5\%$ and an improved $R^2$.

Fig.~\ref{fig:deltatau_vs_fdm_fit} shows that we also get a tight relation between and $M_i^2\Delta \tau(M_i)$  and $f_{DM}$.  %
$\tau$ for DM admixed NS is less than that in the hadronic case. Hence, $\Delta \tau$ is negative and decreases further with more DM fraction. We observe it is roughly a linear fit and fit the following function %
\begin{equation}\label{eqn:deltatau_fit}
    \Delta \tau(M, f_{DM}) =  M^{-2} (C_{\tau} f_{DM}[\%])~. 
\end{equation}
$C_{\tau}$ is the fitting coefficient. $M$ is in units of $M_{\odot}$. Again, we impose ($\Delta \tau(M, f_{DM}=0) =0$) and drop the leading zeroth order term. The fitting coefficient is reported in Table.~\ref{table:deltaf_deltatau_fits}). We can approximate the relation as $\Delta \tau(M, f_{DM}) \approx -0.8 M^{-2}f_{DM}~.$ The bottom panel show that the errors are within 5$\%$ for $f_{DM} \gtrsim 0.01$ and go beyond 20$\%$ for lower DM fractions. Any dependence of the relations Eqs.~(\ref{eqn:deltaf_fit_linear}), (\ref{eqn:deltaf_fit_quadratic}), and (\ref{eqn:deltatau_fit}) on \{nuc\}, if any, is via the fitting parameters. We explore this dependence in Appendix~\ref{sec:appendix_delta_f_tau}, where we conclude that the Eqs.~(\ref{eqn:deltaf_fit_linear}), (\ref{eqn:deltaf_fit_quadratic}), and (\ref{eqn:deltatau_fit}) hold for any hadronic EoS, but the fitting coefficients depends on \{nuc\}.

\begin{table}
    \caption{\label{table:deltaf_deltatau_fits}%
    Fitting coefficients for Eqs.~(\ref{eqn:deltaf_fit_linear}), (\ref{eqn:deltaf_fit_quadratic}), and (\ref{eqn:deltatau_fit}). $C_{fi}$ have the unit kHz/$\sqrt{M_{\odot}}$. $C_{\tau}$ has the units of $M_{\odot}^2$s. $R^2$ is the coefficient of determination, measuring the goodness of each fit.}
    \begin{ruledtabular}
    \begin{tabular}{c|c|cc|c}
       Model& $C_{f1}[10^{-2}]$ & $C_{f2}[10^{-2}]$ & $C_{f3}[10^{-4}]$ & $C_{\tau}[10^{-3}]$\\ \hline
        Hadronic & 1.26 $\pm$ 0.05 & 1.10 $\pm$ 0.14& 1.78 $\pm$ 1.42 & -7.88 $\pm$ 0.49\\ 
        $R^2$ & 0.9942 & \hspace{0.5cm} 0.9994&&0.9991
    \end{tabular} 
    \end{ruledtabular}
\end{table}

\subsection{Effect of Dark Matter II: Variation of all parameters}\label{sec:vary_all}
So far, we kept the nuclear parameters fixed and varied only $G$. We now vary all the parameters (\{nuc\} + $G$) simultaneously and uniformly within their uncertainty ranges. These ranges are given by `Ghosh2022' of Table.~\ref{table:parameters}. We solve for the complex eigen-frequencies for $\sim 6500$ EoSs satisfying $\chi$EFT and `Astro' constraints. 

\begin{figure}
    \centering
    \includegraphics[width=0.95\linewidth,trim={0 0.4cm 0 0},height=0.9\linewidth]{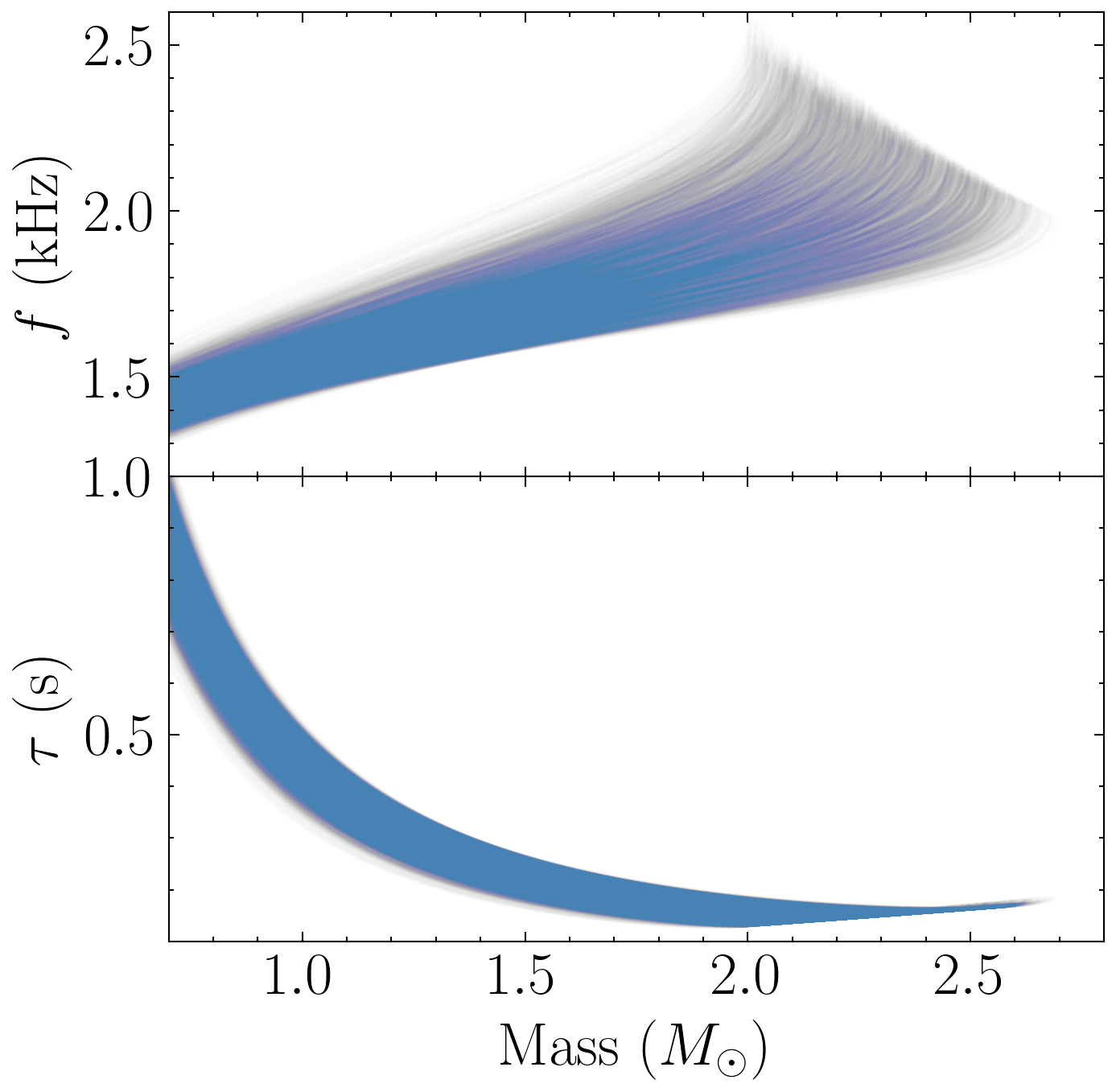}
    \caption{Posteriors of $f$-mode frequency ($f$) and damping time ($\tau$) as a function of NS mass after passing through the $\chi$EFT and `Astro' constraints as outlined in Sec.~\ref{sec:params}. The curves are generated by varying all the microscopic parameters as per the range `Ghosh2022' from Table.~\ref{table:parameters}. Note that all the curves are plotted with the same steel-blue color. The opaqueness varies with the number of overlapping curves. }
    \label{fig:ftau_vs_mass}
\end{figure}

We plot the $f$-mode frequency and the damping times for this posterior ensemble as a function of mass in Fig.~\ref{fig:ftau_vs_mass}. We get a band in the $f-M$ and $\tau-M$ planes. We checked that this overlaps with the band obtained by varying the nuclear parameters without the inclusion of DM. 
This demonstrates that a degeneracy exists between nuclear parameters and DM. The reason for this is that the effect of DM is to soften the EoS, and we impose a $2M_{\odot}$ cut-off which filters out these soft EoSs. The second reason is that in this model, we establish a chemical equilibrium between the neutron and DM particle. Thus, the overall effect is just that of adding an extra degree of freedom throughout the NS. So the band overlaps with one with zero DM. This degeneracy must be considered while constraining the microphysics from future detections of $f$-modes from compact objects and implies that the presence of DM in NS cannot be ruled out. 

For this posterior set, we find that $f$ lies within the range [1.55, 2.0] kHz and [1.67,2.55] kHz for the $1.4M_{\odot}$ and 2$M_{\odot}$ configurations, respectively. The corresponding ranges for $\tau$ are [0.18, 0.30]s and [0.13, 0.20]s respectively. We expect similar ranges for purely hadronic NSs given the degeneracy mentioned above. For completeness, we consider $\sim 3000$ nuclear EoSs with zero DM and vary all the nuclear parameters to check the ranges without DM. For this case, $f$ lies within the range [1.56, 2.0] kHz and [1.68,2.56] kHz for the $1.4M_{\odot}$ and 2$M_{\odot}$ configurations, respectively. The corresponding ranges for $\tau$ are [0.18, 0.29]s and [0.13, 0.19]s respectively. The ranges are similar to those with DM as expected.

\begin{figure}
    \centering
    \includegraphics[width=0.95\linewidth,trim={0 0.5cm 0 0}, height=0.9\linewidth]{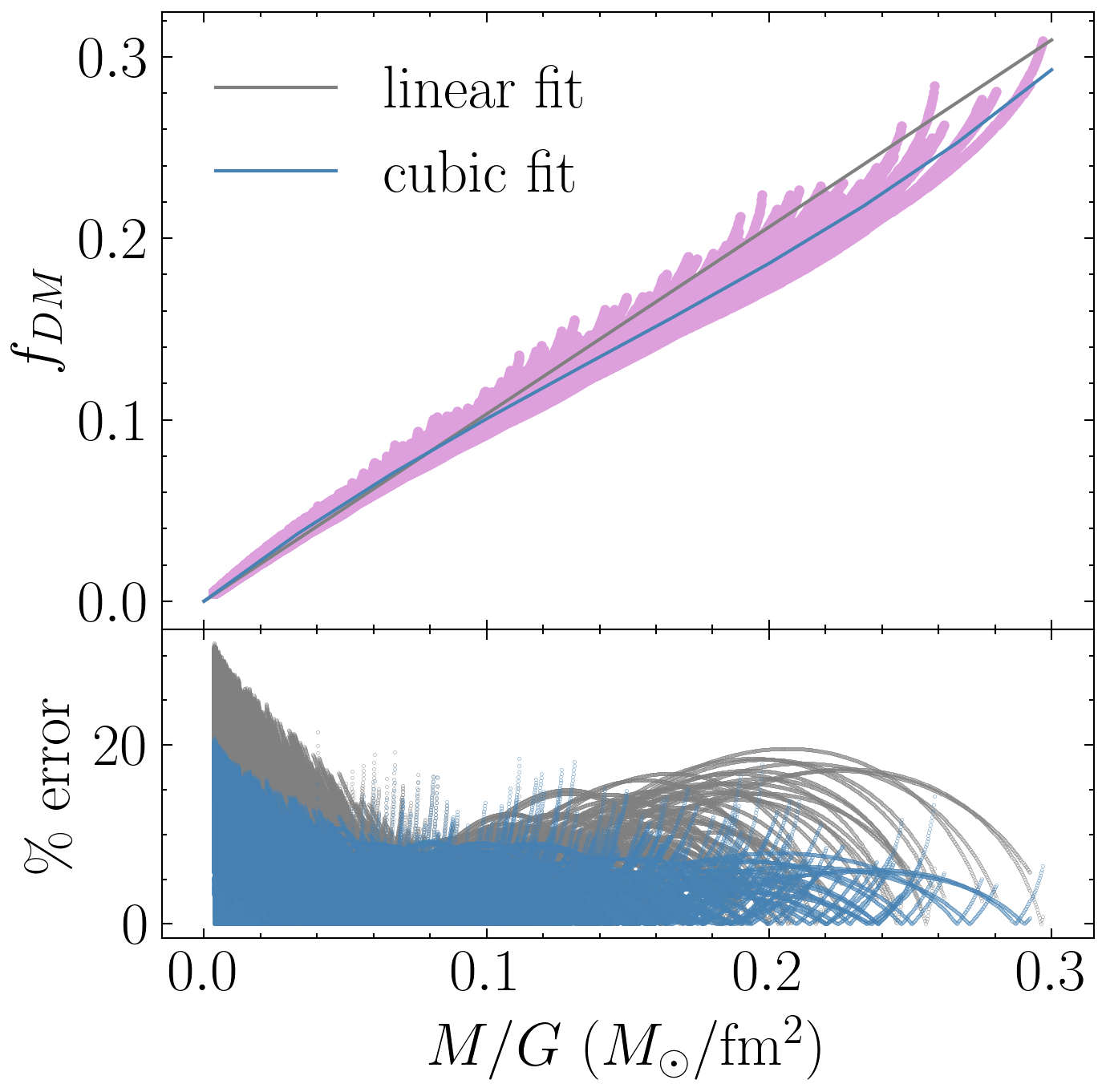}
    \caption{The top panel shows the DM fraction ($f_{DM}$) as a function of $M/G$ obtained by varying all the parameters in the range `Ghosh2022' (refer Table.~\ref{table:parameters}). The gray and blue lines are linear and cubic fits given by Eqs.~(\ref{eqn:dmfrac_fit_linear}) and (\ref{eqn:dmfrac_fit_cubic}), respectively. The fit coefficients are reported in Table.~\ref{table:dmfrac_fits}. The bottom panel shows the percent error for the two fits.}
    \label{fig:dmfrac_fit}
\end{figure}

In the previous section, we noticed the dependence of $\Delta f$ and $\Delta \tau$ on $f_{DM}$. The DM fraction depends on $G$ as well as on the mass of the star. We consider the same posterior sample as generated above. We take 500 mass values in the range [1,2] $M_{\odot}$ and calculate $f_{DM}$ for each mass configuration $M_i$ for all the EoSs. We find a linear dependence of $f_{DM}$ on $1/G$ for each mass value $M_i$, with larger slopes for larger $M_i$. Analyzing the data, we find that the slope increases roughly linearly with mass. Hence, we make a plot of $f_{DM}$ as a function of $M/G$ (See Fig.~\ref{fig:dmfrac_fit}) and get a fairly good relation.%

We perform linear and cubic fits of the form%
\begin{align}
    f_{DM} &= C_1 \left(\frac{M}{G}\right)~, \label{eqn:dmfrac_fit_linear}\\
    f_{DM} &=  C_2 \left(\frac{M}{G}\right) + C_3 \left(\frac{M}{G}\right)^2 + C_4 \left(\frac{M}{G}\right)^3~. \label{eqn:dmfrac_fit_cubic}
\end{align}
$C_i$ are the fitting coefficients. $M$ and $G$ are in units of $M_{\odot}$ and fm$^2$ respectively. Note that we have varied all the microscopic parameters here, making the relation obtained for $f_{DM}(M, G)$ universal. Given a DM self-interaction strength value, the DM fraction in a DM admixed NS of a given mass configuration is independent of the hadronic EoS. We recover purely hadronic NS as asymptotically large values of $G$, i.e., $\lim_{G\to\infty} f_{DM}= 0$. This fixes the leading zeroth order term to be zero. 

The fit coefficients are reported in Table~\ref{table:dmfrac_fits}%
\begin{table}
    \caption{\label{table:dmfrac_fits}%
    Values of fitting coefficients for Eqs.~(\ref{eqn:dmfrac_fit_linear}) and (\ref{eqn:dmfrac_fit_cubic}). $R^2$ is the coefficient of determination, measuring the goodness of each fit.}
    \begin{ruledtabular}
    \begin{tabular}{c|c|ccc}
       Model& $C_{1}$ $\left[\frac{\text{fm}^2}{M_{\odot}}\right]$ & $C_{2}$ $\left[\frac{\text{fm}^2}{M_{\odot}}\right]$ & $C_{3}$ $\left[\frac{\text{fm}^4}{M_{\odot}^2}\right]$ & $C_{4}$ $\left[\frac{\text{fm}^6}{M_{\odot}^3}\right]$\\ \hline
        Ghosh22 \cite{ghosh2022multi} & 1.03  & 1.20  & -3 & 6 \\
        &$\pm0.02$&$\pm 0.05$&$\pm1$&$\pm4$\\ \hline
        $R^2$ & 0.9876 && 0.9972& \\
    \end{tabular} 
    \end{ruledtabular}
\end{table}
The linear relation fits to an accuracy of 30$\%$. For $M/G \gtrsim 0.04 M_{\odot}^2/\text{fm}^2$ the fit is within 20$\%$. The cubic relation stays within an error of 20$\%$. Since we vary all the parameters, we encounter higher DM fractions (up to 30$\%$). This is in line with the upper limit on $f_{DM}$ of $37.9\%$ found in our previous study~\cite{Shirke2023b}. These are the EoS with stiff hadronic EoS with low G, i.e., with a high amount of DM. The cases with larger $f_{DM}$ are filtered out as they lead to very soft EoS violating the 2$M_{\odot}$ pulsar mass constraint. This filter results in fewer points on the right side of this plot. The coefficient of the linear fit is close to one. Adopting the linear relation, we get an approximate relation as 
\begin{equation}\label{eqn:dffrac_final_relation}
    f_{DM} \approx \left(\frac{M}{M_{\odot}}\right)\left(\frac{\text{fm}^2}{G}\right)~,
\end{equation}
which can used as a quick estimator of the dark matter fraction. We can also use this in Eqs.~(\ref{eqn:deltaf_fit_linear}), (\ref{eqn:deltaf_fit_quadratic}), and (\ref{eqn:deltatau_fit}) to determine the change in $f$-mode frequency and damping time in terms of mass configuration and self-interaction strength.

\subsection{Correlation Studies}\label{sec:correlation}
Having studied the effect of DM, we now perform a correlation study to check the effect of microscopic parameters on NS observable, particularly the $f$-mode parameters. We consider the nuclear parameters \{nuc\} and DM interaction strength $G$ for the microscopic parameters. For NS observables, we consider the maximum mass ($M_{max}$), the radius, and the tidal deformability of $1.4 M_{\odot}$ star ($R_{{1.4M_{\odot}}}$; $\Lambda_{{1.4M_{\odot}}}$) and $2 M_{\odot}$ star ($R_{{2M_{\odot}}}$; $\Lambda_{{2M_{\odot}}}$). Also, for the $f$-mode observables, we consider the frequency and damping time of $1.4 M_{\odot}$ star ($f_{{1.4M_{\odot}}}$; $\tau_{{1.4M_{\odot}}}$) and $2 M_{\odot}$ star ($f_{{2M_{\odot}}}$; $\tau_{{2M_{\odot}}}$). We also consider the corresponding DM fractions ($f_{DM,1.4M_{\odot}}; f_{DM,2M_{\odot}}$). The correlation between any two parameters ($x$, $y$) is calculated using Pearson's coefficient for linear correlation ($r(x,y)$) given by
\begin{align}
    &r(x,y) = \frac{cov(x,y)}{\sqrt{cov(x,x)cov(y,y)}}~, \\ 
    &\text{where, } 
    cov(x,y) = \frac{1}{N}\sum_{i=1}^{N}(x_i-\bar{x})(y_i-\Bar{y})~.
\end{align}
We study the correlations by varying all the nuclear parameters in the range `Ghosh2022'. We also check how correlations change if the effective mass $m^*/m$ is precisely known as it is the most dominant parameter. For each case, we apply all the $\chi EFT$, $2M_{\odot}$ pulsar mass, and the tidal deformability constraint from GW170817. 

\subsubsection{Variation of all parameters}\label{sec:correlation_all}
\begin{figure*}
    \includegraphics[width=0.81\linewidth,trim={0 0.0cm 0 0.0cm}]{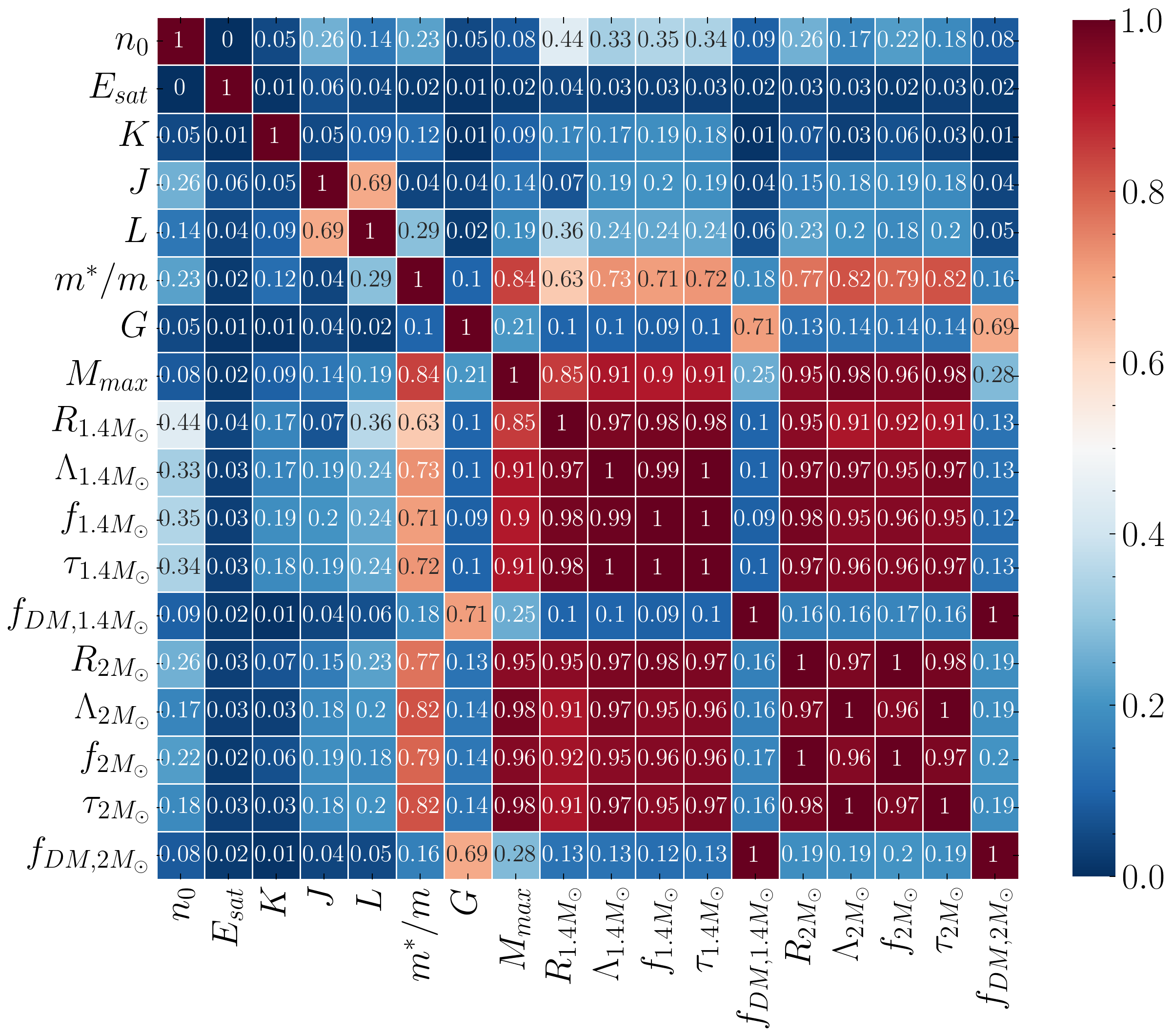}
    \caption{Correlation matrix showing the correlations among the nuclear parameters, DM interaction parameter, NS observables and the f-mode characteristics. Correlations are obtained after applying the $\chi EFT$, GW170817, and $2M_{\odot}$ pulsar mass constraints. The parameter range is given in Table.~\ref{table:parameters} }
    \label{fig:Correlation_allvary}
\end{figure*}

The variation range of the nuclear and DM parameters are given in Table~\ref{table:parameters} labelled by `Ghosh2022'. 
The correlation matrix among the \{nuc\}, $G$, and NS properties resulting after consideration of $\chi EFT$, $2M_{\odot}$ pulsar mass, and GW170817 constraints is displayed in Fig.~\ref{fig:Correlation_allvary}.%
We make the following observations:

\begin{itemize}
    \item We find a strong correlation between $J$ and $L$ (0.69). This is expected due to $\chi EFT$ constraints and is consistent with previous studies.
    \item The effective mass $m^*/m$ shows a strong correlation with the NS properties and $f$-mode characteristics for both $1.4M_{\odot}$ and $2M_{\odot}$ stars. The correlation of  $m^*/m$ with the NS properties is consistent with previous studies~\cite{ghosh2022multi, Ghosh2022b, Shirke2023a}.  $F$-mode characteristics in turn depend on these macroscopic properties (see Sec.~\ref{sec:UR}), which explains the strong correlation of $m^*/m$ with $f$-mode parameters. The saturation density $n_0$ shows a moderate correlation with $1.4M_{\odot}$ properties.
    \item All NS observables are strongly correlated with each other as well as with $f$-mode observables.
    \item The DM fraction for both $1.4M_{\odot}$ and $2M_{\odot}$ stars strongly correlates with the DM self-interaction parameter $G$ (0.71 and 0.69, respectively). This is consistent with our previous finding of Eq.~(\ref{eqn:dffrac_final_relation}) that $f_{DM} \sim 1/G$. Further, $f_{DM,1.4M_{\odot}}$ and $f_{DM,2M_{\odot}}$ are perfectly correlated, which also follows from  Eq.~(\ref{eqn:dffrac_final_relation}) ($f_{DM,2M_{\odot}} = 2f_{DM,1.4M_{\odot}}/1.4$) since $G$ is fixed for a given EoS.
    \item $G$, $f_{DM,1.4M_{\odot}}$, $f_{DM,2M_{\odot}}$ do not show correlations with any other parameters. 
\end{itemize}

The posterior distribution of the dominant parameters is discussed in Appendix~\ref{sec:appendix_posterior}. We find that 90$\%$ quantiles for $f_{DM, 1.4M_{\odot}}$ and $f_{DM, 2M_{\odot}}$ are $3.97\%$ and $5.79\%$, respectively. Thus, the model prefers only low DM fractions. In Appendix~\ref{sec:appendix_posterior}, we also discuss how the posteriors are affected when a filter of higher pulsar mass of $M=2.3M_{\odot}$ is used. This is motivated by the recent observation of a heavy black widow pulsar PSR J0952-0607~\cite{Romani2022}. We find that the existence of a NS with mass as high as $2.3M_{\odot}$ restricts DM fraction to even lower values. The 90$\%$ quantiles for $f_{DM, 1.4M_{\odot}}$ and $f_{DM, 2M_{\odot}}$ reduce to $3.03\%$ and $4.29\%$, respectively. Thus, heavy NSs disfavor the presence of DM in NSs. This is because the presence of DM softens the EoS, and the higher masses filter-out soft EoSs.

We also check the effect of fixing $E_{\rm sat}$ and $J$ to $-16$ MeV and $31$ MeV, respectively (not shown). This is checked as these parameters are well-constrained from experiments. This leads to a moderate correlation of $L$ with NS observables and $m^*/m$. The effective mass $m^*/m$ remains the dominant parameter dictating the NS macroscopic properties. We infer from this study that the NS and $f$-mode observables are affected mainly by the nuclear parameter $m^*/m$. Given the uncertainty of the nuclear parameters, we do not find strong correlations of any observables with the DM interaction strength $G$. 

\subsubsection{Fixing $m^*/m$}
\begin{figure}
    \centering
    \includegraphics[width=\linewidth,trim={0 0.4cm 0 0}]{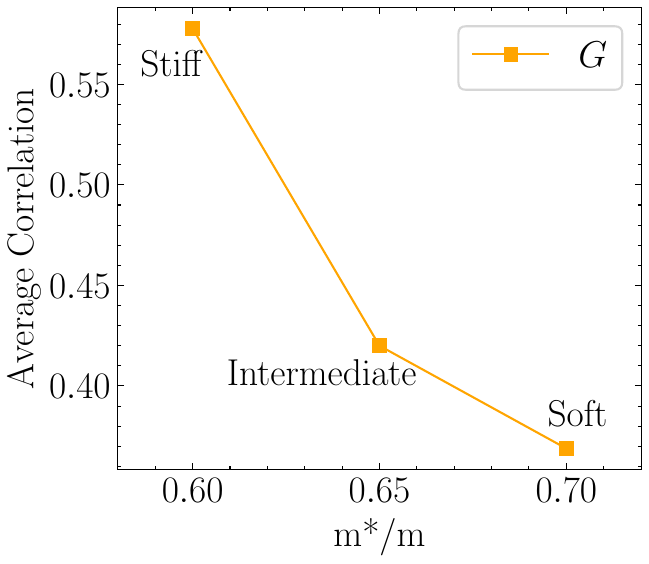}
    \caption{The average correlation of $G$ with NS observables is plotted when the effective mass ($m^*/m$) is fixed to different values. We consider the values 0.60, 0.65, and 0.70 for the effective mass, corresponding to a stiff, intermediate, and soft EoS, respectively.}
    \label{fig:average_correlation}
\end{figure}

It is observed that $m^*/m$ has the strongest correlations with the NS observables. We check the effect on correlation in case future experiments measure the nuclear equation of state at high densities, i.e.\ the effective mass parameter in our approach, precisely. This could help constrain the DM self-interaction parameter $G$ better. We consider three different values for the effective mass: 0.6, 0.65, and 0.7, capturing the stiff, intermediate, and soft cases of EoS, respectively. Here, we focus on how the overall correlation of $G$ is affected when $m^*/m$ is fixed to different values. For this, we take the average of the correlation of $G$ with all the observables mentioned above. The detailed correlations for these cases are displayed in Appendix~\ref{sec:appendix_fixed_mstar}. We define ``Average Correlation" as the arithmetic mean of correlation of $G$ with all the observables namely, $R$, $\Lambda$, $f$, $\tau$ of 1.4$M_{\odot}$ and 2$M_{\odot}$ configurations and $M_{max}$. We plot this average correlation as a function of $m^*/m$ in Fig.~\ref{fig:average_correlation}. Note, these numbers are only to see the dominance of $G$ in dictating the NS observables when $m^*/m$ is fixed.

We find that when $m^*/m$ is fixed to 0.60, the average correlation of $G$ is 0.58. The correlation reduces to 0.42 and 0.37 as $m^*/m$ is increased to 0.65 and 0.70, respectively. Thus, the correlation decreases with an increased fixed value of $m^*/m$. Lower $m^*/m$ corresponds to stiffer EoS. This leads to a hadronic EoS with a large maximum mass. This makes it possible for DM to soften the EoS, making it an important parameter to dictate the maximum mass and other observables. So we get more distinguishing power for stiffer EoSs compared to the softer ones. However, when $m^*/m$ is high, the hadronic NS has a lower maximum mass to begin with. As DM is known to reduce the maximum mass, we can have only a restricted amount of allowed DM (corresponding to a restricted range of $G$), keeping the total mass above 2$M_{\odot}$. It is because of this restriction in range imposed by the maximum mass condition that the relative importance of the $G$ reduces with increasing $m^*/m$. The detailed comparison of each correlation of $G$ and the other nuclear parameters when $m^*/m$ is fixed can be found in Appendix~\ref{sec:appendix_fixed_mstar}.

\subsection{Universal Relations}\label{sec:UR}

\begin{figure}
    \centering
    \includegraphics[width=\linewidth,trim={0 0.4cm 0 0}]{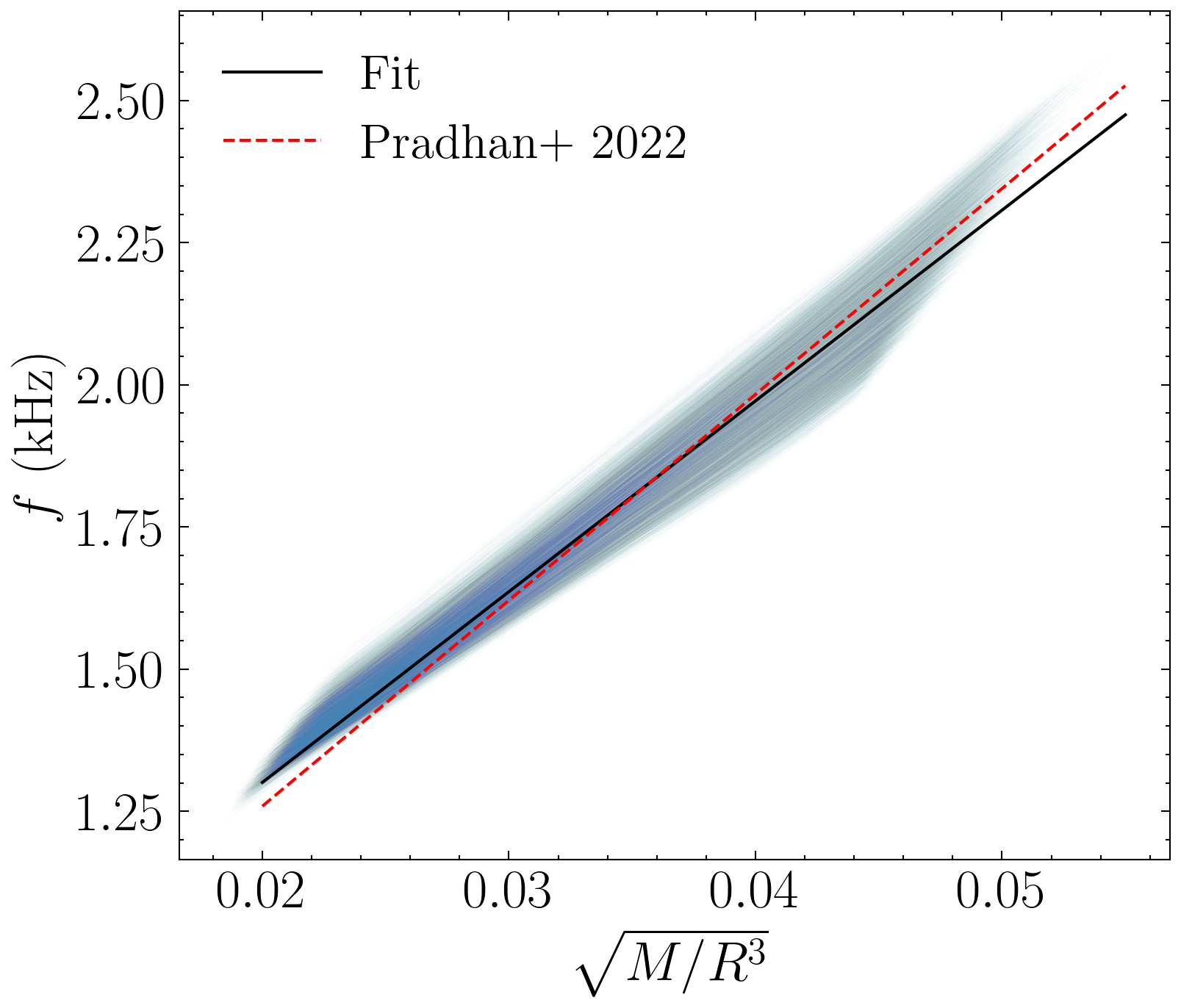}
    \caption{$f$-mode frequency as a function of average density.}
    \label{fig:f_vs_dens}
\end{figure}

We check some universal relations involving $f$-mode frequency and damping time. It was shown by Andersson and Kokkotas~\cite{AnderssonKokkotas1996, AnderssonKokkotas1998} that the $f$-mode frequency is a function of the average density. The relation between the $f$-mode frequency and density is of the form,
\begin{equation}\label{eqn:f_vs_dens}
    f\text{ (kHz)} = a + b\sqrt{\frac{M}{R^3}}~,
\end{equation}
where $M$ is the total mass of the star, and $R$ is its radius. The parameters $a$ and $b$ give the best-fit coefficients. Such fits were obtained by~\cite{DOneva2013, Pradhan2022} for $f$-modes calculated in full GR. We plot $f$ as a function of the square root of the average density in Fig.~\ref{fig:f_vs_dens}. We get a linear relation as expected. We plot the previously obtained best-fit line~\cite{Pradhan2022} with $a=0.535$ kHz and $b=36.20$ kHz-km. This relation between $f$ and $M/R^3$ is rather model-dependent, and we do not get a tight relation. We perform our own fit as our model includes DM. The fitting coefficients obtained are $a=0.630$ kHz and $b=333.544$ kHz-km. These coefficients are tabulated in Table~\ref{table:f_vs_dens}. We also include results from other previous work in the table that derived the fitting coefficients for $f$-modes calculated in full GR. The fit in this work corresponds to the case of DM admixed NS $f$-modes in a full-GR setup. A previous work~\cite{Das2021} also performed this kind of fit for DM admixed NS but for a different DM model within the Cowling approximation.

\begin{table}
    \caption{\label{table:f_vs_dens}%
    Values of fitting coefficients for Eq.~(\ref{eqn:f_vs_dens}) from different works.}
    \begin{ruledtabular}
    \begin{tabular}{c|cc}
       Work& $a$ (kHz)& $b$ (kHz km)\\ \hline
       Andersson and Kokkotas~\cite{AnderssonKokkotas1998} & 0.22 &  47.51 \\
       Benhar \& Ferrari~\cite{BenharFerrariGualtieri2004} & 0.79 & 33 \\
       Pradhan+ 2022~\cite{Pradhan2022} & 0.535 & 36.2 \\
        This work & 0.630  & 33.54\\
    \end{tabular} 
    \end{ruledtabular}
\end{table}

\begin{figure}
    \includegraphics[width=\linewidth,trim={0 0.5cm 0 0}, height=0.7\linewidth]{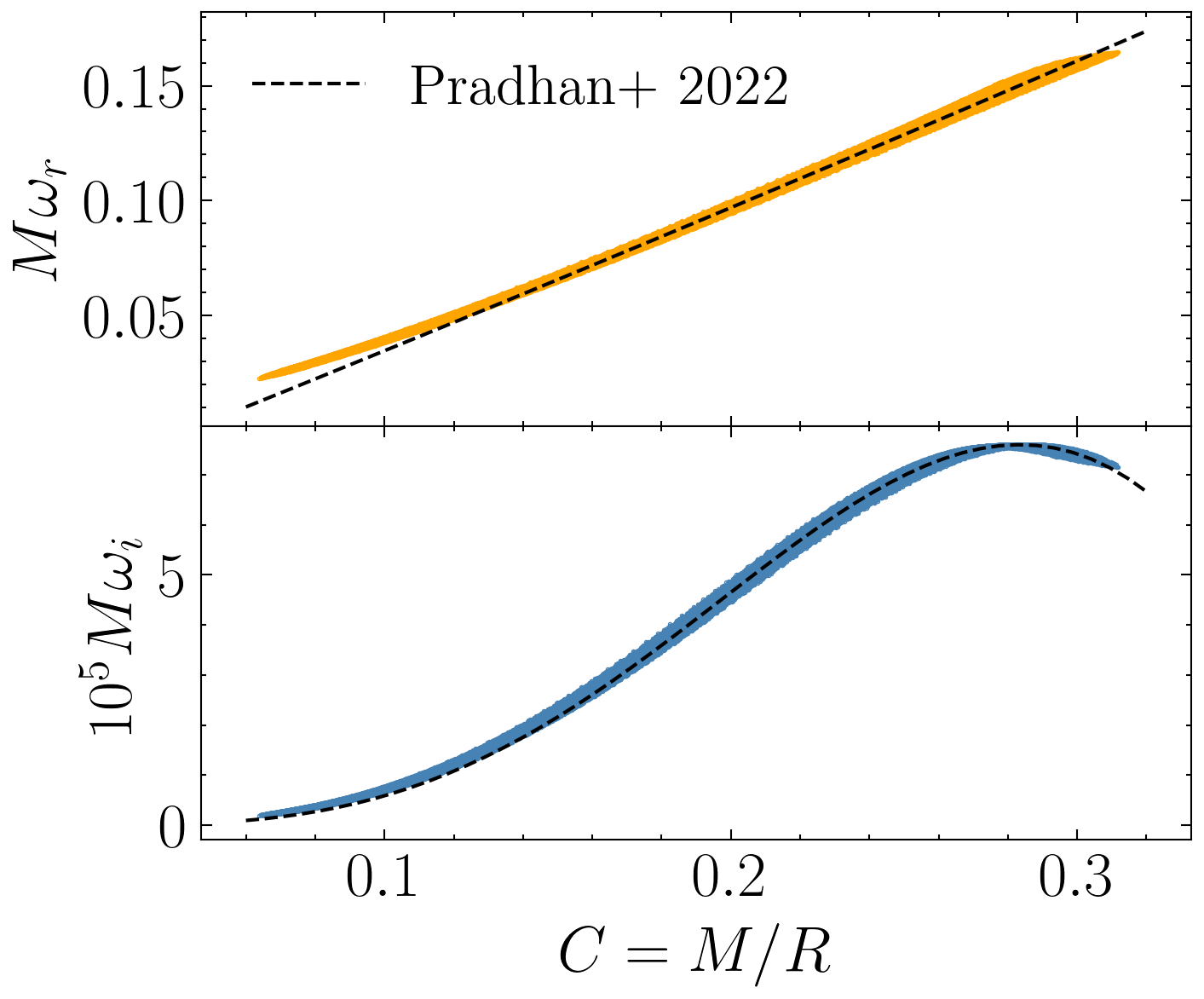}
    \caption{Mass-scaled complex $f$-mode frequency as a function of compactness. The upper panel shows the real part ($M\omega_r$) representing the mass-scaled frequency and the lower panel shows the imaginary part ($M\omega_i$) denoting the damping time ($M\omega_i = M/\tau$).}
    \label{fig:UR_MOmega_vs_C}
\end{figure}

There are other relations that are model-independent that we call universal relations. It was shown in Ref.~\cite{AnderssonKokkotas1998} that both components of the complex eigen-frequency ($\omega = \omega_r + i\omega_i$) when scaled with mass (M) show a tight correlation with compactness. Here, $\omega_r = 2 \pi f$ is the $f$-mode angular frequency, and $\omega_i=1/\tau$ is the inverse of damping time. These universal relations are of the form
\begin{align}\label{eqn:UR_C}
    M\omega_r &= a_0 + a_1 C + a_2C^2~, \nonumber \\ 
    M\omega_i &= b_0C^4 + b_1C^5 + b_2C^6~.
\end{align}
Here $C=M/R$ is the dimensionless compactness. The parameters $a_i$ and $b_i$ are obtained by performing a best-fit analysis. Ref.~\cite{Pradhan2022} obtained such a fit using a large set of nuclear EoS and hyperonic EoS. 
We plot $M\omega_r$ and $M\omega_i$ as a function of compactness in~\cref{fig:UR_MOmega_vs_C}. We also plot the best-fit relation as obtained in \cite{Pradhan2022} and find the DM admixed NSs agree with the universal relation. 

\begin{figure}
    \includegraphics[width=\linewidth,trim={0 0.5cm 0 0}, height=0.7\linewidth]{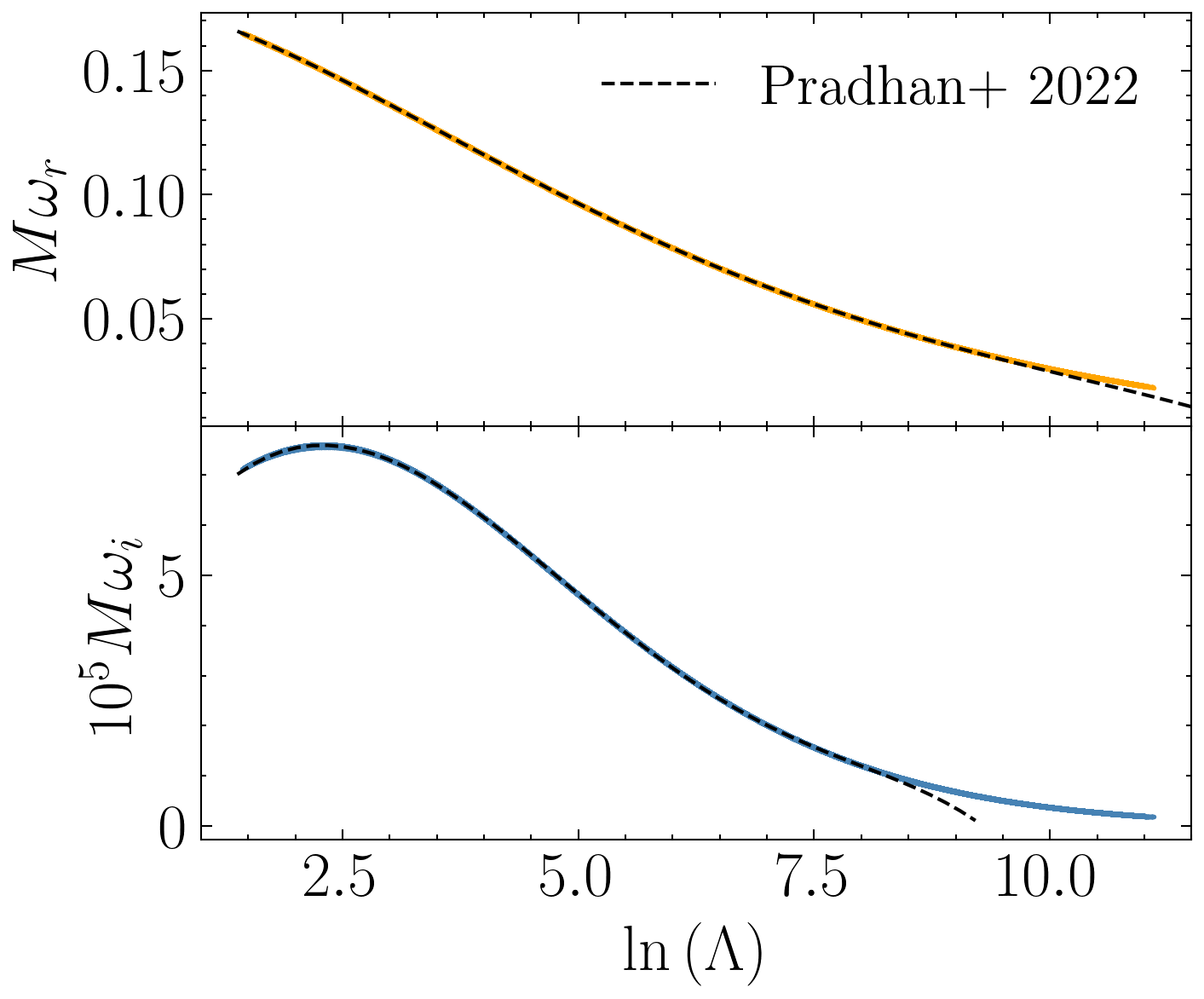}
    \caption{Mass-scaled complex $f$-mode frequency as a function of tidal deformability. The upper panel shows the real part ($M\omega_r$) representing the mass-scaled frequency and the lower panel shows the imaginary part  ($M\omega_i$) denoting the mass-scaled damping time ($M\omega_i = M/\tau$).}
    \label{fig:UR_MOmega_vs_lam}
\end{figure}
Another universal relation exists between the mass-scaled complex $f$-mode frequency and the dimensionless tidal deformability~\cite{Chan2014, SotaniKumar2021, Pradhan_2023PRD}. This is given by
\begin{equation}\label{eqn:UR_lam}
    M\omega = \sum_{i} \alpha_i (\ln{\Lambda})^i~.
\end{equation}
We plot $M\omega_r$ and $M\omega_r$ as a function of $\ln{\Lambda}$ in Fig.~\ref{fig:UR_MOmega_vs_lam}. The DM admixed NS considered here are found to follow these relations. The fitting coefficients obtained in~\cite{Pradhan2022} are only for $\ln{\Lambda} \lesssim 8$. Here, we plot it for higher values, where the fit appears to diverge from the universal relation. We note that this relation is the most tight universal relation among all cases studied..

There also exists a universal relation between the mass-scaled frequency and mass scaled damping time (Eq.~(\ref{eqn:UR_f_tau})) which was already explored in Sec.~\ref{sec:vary_G}. The red curves in Fig.~\ref{fig:f_vs_tau_diffm} are obtained using this relation for $1.2M_{\odot}$, $1.4M_{\odot}$, $1.6M_{\odot}$, $1.8M_{\odot}$, and $2.0M_{\odot}$ configurations where the fit coefficients used are provided in Table VI of~\cite{Pradhan2022}. From this, we infer that the NS admixed DM $f$-mode characteristics also obey this universal relation.

We conclude that for the DM model considered here, DM admixed NSs follow the existing universal relations of $f$-modes. It is, therefore, evident that in $f$-mode detections, DM admixed NSs can masquerade as purely hadronic neutron stars, and one needs to look beyond GR effects to lift the degeneracy.


\section{\label{sec:discussions}Discussions}

In this work, we perform a systematic investigation of the non-radial quadrupolar fundamental modes of oscillations of DM admixed NSs within the full general relativistic framework. For the hadronic matter EoS, we use the phenomenological relativistic mean field model with nucleons strongly interacting via the exchange of mesons. We consider the model based on neutron decay anomaly for DM, which allows for a large DM fraction within NSs. Assuming a chemical equilibrium between the neutron and the DM particle, we solve for the structure equations, $f$-mode oscillation frequency, and damping time for DM admixed NSs in a single-fluid formalism. We only consider those hadronic microscopic parameters that are consistent with the chiral effective field theory calculations at low densities and follow the astrophysical constraints from the present electromagnetic and gravitational wave data. 

We first studied the effect of the inclusion of DM on $f$-mode oscillation of NSs. We fixed the hadronic EoS and found that the $f$-mode frequency for a given mass configuration increases when we include DM within the NS. The effect is similar when we consider configurations of fixed compactness and tidal deformability. The change in the $f$-mode characteristic is higher for high mass, high compactness, and low tidal deformability configurations. The effect is similar to that of a softer EoS since we know that the inclusion of DM softens the EoS. The opposite effect is seen for the damping time ($\tau$), where $\tau$ reduces upon the inclusion of DM. Similar to frequency, the change in damping time is higher for high mass, compactness, and low tidal deformability configurations. 

We then checked the effect of DM self-interaction strength ($G$) and DM fraction ($f_{DM}$) on $f$-mode characteristics. As  $G$ is varied, we found that $f$ ($\tau)$ decreases (increases) with an increase in $G$. The opposite effect is seen when considering $f_{DM}$. This is expected, as we know $f_{DM}$ to be less for larger $G$, which results in a stiffer EoS. Larger $G$ increases the energy cost to create DM particles, resulting in less DM fraction and stiffer EoS, which is closer to the purely hadronic case. When $G$ is varied, and the resulting characteristics are plotted on the $f-\tau$ plane, it is seen to follow the universal mass-scaled $f-\tau$ relation. We found that, in contrast to $G$, $f$ and $\tau$ vary linearly with $f_{DM}$. 

We explore this dependence in detail. We defined a quantity $\Delta f$ and $\Delta \tau$ where we subtract out the effect from the purely hadronic part ($f_{DM}=0$). Analyzing these quantities, we derived a relation for them in terms of the DM fraction $f_{DM}$ and the mass configuration. We found that $\Delta f \sim \sqrt{M}(af_{DM}+bf_{DM}^2)$, and $\Delta \tau \sim -f_{DM}/M^2$. These are new relations and directly tell the effect on the change in the $f$-mode frequency and damping time for any mass configuration and DM fraction. These relations hold for any hadronic EoS, only the coefficients change.

We then systematically varied all the nuclear and DM parameters simultaneously. Correspondingly, we got a band in the $f-M$ and $\tau-M$ plane. This band is the same as that we get just from the variation of nuclear parameters without DM, demonstrating the degeneracy between NS and DM admixed NS, which needs to be considered while constraining models from $f$-mode observations.
The range of $f_{1.4M_{\odot}}$ and $f_{2M_{\odot}}$ is [1.55,2.0] kHz and [1.67,2.55] kHz, respectively, and that for $\tau_{1.4M_{\odot}}$ and $\tau_{2M_{\odot}}$ is [0.18, 0.30] s and [0.13, 0.20] s respectively. We further found a relation between the DM fraction ($f_{DM}$), the star's gravitational mass ($M$), and the self-interaction parameter ($G$) as $f_{DM} = 1.03(M/M_{\odot})(\text{fm}^2/G)$. This relation is universal and predicts the DM fraction of a DM admixed NS of mass $M$ with DM self-interacting with strength $G$.

For this set of EoSs, we also checked for physical correlations. Keeping only those EoSs consistent with the $\chi$EFT calculations at low densities and also satisfying the astrophysical constraints of $2M_{\odot}$ and the tidal deformability from GW170817, we looked for any physical correlations between microscopic nuclear and DM parameters and NS macroscopic observables. Among the DM parameters, we found a strong correlation only between $f_{DM}$ and $G$. This is consistent with our previous analysis, where we obtained $f_{DM} \sim 1/G$. From the posterior distribution, we obtain the 90$\%$ quantiles for $f_{DM, 1.4M_{\odot}}$ and $f_{DM, 2M_{\odot}}$ are $3.97\%$ and $5.79\%$, respectively. Thus, only low DM fractions are favored. Our analysis shows that observations of heavy NSs disfavor the presence of DM for the considered DM model and may rule out the presence of DM. The effective mass is the most dominant parameter to dictate the macroscopic properties. For this reason, we checked the correlations in case $m^*/m$, i.e. the nuclear equation of state, is precisely measured in future experiments. Upon fixing $m^*/m$ to 0.6, we found an emergence of a strong correlation for $n_0$ and $G$. As we increase the value of $m^*/m$ to 0.65 and 0.7 (stiff to soft EoS), the correlation of $G$ weakens and that of $n_0$ and other nuclear parameters, $K$, and $L$ strengthens. In such a case, $G$ becomes the next dominant parameter to dictate the maximum mass.

Finally, we used this set of EoS to check the universal relations of $f$-modes. We fitted a linear relation between $f$ and the square root of average density ($M/R^3$) and reported the fitting coefficients for the DM admixed NSs. We then checked the universal relations of the $f$-mode characteristics with compactness and tidal deformability. These are found to follow the previously known relations for neutron stars without DM and reiterate the degeneracy between NS models and DM admixed NS models. 


This work explores the $f$-mode characteristics of DM admixed NS in a full-GR setup for the DM model considered. A parallel study~\cite{Flores2024} calculating $f$-modes for DM admixed NS appeared during the completion of this work. They adopt a different model where the DM interactions are mediated via the Higgs boson. The effect of DM on $f$-mode frequency and damping time is consistent with what we observe, i.e., $f$ ($\tau$) increases (decreases) with an increase in the amount of DM. One previous work~\cite{Das2021} that studied these $f$-modes adopted Cowling approximation and used select equations of state. They also employ a different DM model. Another work~\cite{GleasonBrownKain2022} that studied oscillations of DM admixed NS in full relativistic setup simulates the evolution dynamically and focuses only on the radial pulsations.
In summary, the results presented in our investigation are important in light of future BNS merger events expected from upcoming GW observations, which will enable tighter constraints on $f$-mode frequencies and their role in delineating the constraints on DM models.

\begin{acknowledgments}\label{sec:acknowledgements}
S.S. and B.K.P. acknowledge the use of the Pegasus Cluster of IUCAA's high-performance computing (HPC) facility, where numerical computations were carried out. L.S. and J.S.B. acknowledge support by the Deutsche Forschungsgemeinschaft (DFG, German Research Foundation) through
the CRC-TR 211 `Strong-interaction matter under extreme conditions' --
project no.\ 315477589 -- TRR 211.
\end{acknowledgments}
\appendix
\input{appendix}
\bibliography{Refs}
\end{document}

%% file: appendix.tex
\section{Differential Equations for solving the Non-radial Quasi normal modes (QNM) of Compact stars }\label{sec:appendix_qnm_eqs}
Here, we present the basic equations that need to be solved for finding the complex QNM frequencies.
\subsubsection{Perturbations Inside the Star}
The perturbed metric ($ds^2_p$) can be written as ~\cite{Thorne},
\begin{eqnarray}
    ds^2_p=ds^2+h_{\mu \nu} dx^{\mu}dx^{\nu}~.
    \label{eqn:perturbedmetric}
\end{eqnarray}

Following the arguments given in Thorne and Campolattaro~\cite{Thorne}, we focus on the even-parity (polar) perturbations for which the the GW and matter perturbations are coupled. Then $h_{\mu \nu}$ can be expressed  as ~\cite{Sotani2001,Thorne},
\begin{eqnarray}
    h_{\mu \nu}=
   \begin{pmatrix}
r^lHe^{2\Phi} & i\omega r^{l+1} H_1 &0&0\\
i\omega r^{l+1} H_1 & r^l H e^{2\lambda} &0 &0\\
0 &0 & r^{l+2}K& 0\\
0 & 0& 0 &  r^{l+2}K sin^2{\theta}
\end{pmatrix} Y^l_m e^{i\omega t} ~, \nonumber\\
 \text{ }
 \label{eqn:metricfunctions}
\end{eqnarray}
where $Y_m^l$ are spherical harmonics. $H,\ H_1, \ K$ are  perturbed metric functions and vary with $r$ (i.e.,  $H=H(r),\ H_1=H_1(r), \ K=K(r)$ ). The  Lagrangian displacement vector  $\textbf{$\zeta$}=(\zeta^r,\zeta^{\theta},\zeta^{\phi})$  associated with the polar perturbations of the fluid can be characterized as,

\begin{eqnarray}
    \zeta^{r}&=&\frac{r^l}{r}e^{-\lambda} W(r)  Y^l_m e^{i\omega t} \nonumber \\
    \zeta^{\theta}&=&\frac{-r^l}{r^2} V(r)  \frac{\partial Y^l_m}{\partial \theta} e^{i\omega t} \nonumber \\
    \zeta^{\phi}&=&\frac{-r^l}{r^2 sin^2\theta} V(r)  \frac{\partial Y^l_m}{\partial \phi} e^{i\omega t}
    \label{eqn:pertfluid}
\end{eqnarray}

where $W,V$ are amplitudes of the radial and transverse fluid perturbations. The equations governing these perturbation functions and the metric perturbations inside the star are given by~\cite{Sotani2001},
\begin{eqnarray}
    \frac{d H_1}{dr}&=&\frac{-1}{r}\left[l+1+\frac{2m}{r}e^{2\lambda}+4\pi r^2e^{2\lambda} \left( p-\epsilon \right)\right] H_1 \nonumber\\
    &+&\frac{1}{r}e^{2\lambda}\left[H+K+16\pi\left(p+\epsilon\right)V\right] \label{eqn:dh1} \ , \\
    \frac{d K}{dr}&=&\frac{l\l(l+1\r)}{2r}H_1+\frac{1}{r}H-\l(\frac{l+1}{r}-\frac{d\Phi}{dr}\r)K \nonumber \\
    &+&\frac{8\pi}{r}\l(p+\epsilon\r)e^{\lambda} W \ ,  \label{eqn:dk} \\
    \frac{d W}{dr}&=&re^{\lambda}\l[\frac{1}{\gamma p}e^{-\Phi}X-\frac{l\l(l+1\r)}{r^2}V-\frac{1}{2}H-K\r] \nonumber \\
    &-&\frac{l+1}{r}W   \label{eqn:dw} \ , \\
    \frac{d X}{dr}&=& \frac{-l}{r}X+\l(p+\epsilon \r)e^{\Phi}\Bigg[\frac{1}{2}\l(\frac{d\Phi}{dr}-\frac{1}{r}\r)H \nonumber\\
    &-&\frac{1}{2}\l( \omega^2re^{-2\Phi}+\frac{l(l+1)}{2r}\r)H_1+\l(\frac{1}{2r}-\frac{3}{2}\frac{d\Phi}{dr}\r)K \nonumber\\
    &-&\frac{1}{r}\l[ \omega^2\frac{e^{\lambda}}{e^{2\Phi}}+4\pi \l(p+\epsilon \r) e^{\lambda}-r^2\frac{d}{dr}\l( \frac{e^{-\lambda}}{r^2}\frac{d \Phi}{dr}\r)\r]W \nonumber \\
    &-&\frac{l(l+1)}{r^2}\frac{d\Phi}{dr}V\Bigg]  \label{eqn:dx}\ ,
\end{eqnarray}
\begin{eqnarray}
   &&\l[1-\frac{3m}{r}-\frac{l(l+1)}{2}-4\pi r^2p\r]H-8\pi r^2 e^{-\Phi}X \nonumber\\
   &-& \l[ 1+ \omega^2r^2e^{-2\Phi} -\frac{l(l+1)}{2}-(r-3m-4\pi r^3p)\frac{d\Phi}{dr}\r]K \nonumber \\
   &+&r^2e^{-2\lambda}\l[\omega^2e^{-2\Phi}-\frac{l(l+1)}{2r}\frac{d\Phi}{dr}\r]H_1 =0  \label{eqn:h}\\
  && e^{2\Phi}\l[ e^{-\phi}X+\frac{e^{-\lambda}}{r}\frac{dp}{dr} W+\frac{(p+\epsilon)}{2}H\r] \nonumber \\
  &-&\omega^2 \l(p+\epsilon\r) V=0 ~, \label{eqn:v}
\end{eqnarray}

where $X$ is introduced as~\cite{Detweiler83,Sotani2001}
\begin{eqnarray}
    X&=&\omega^2\l(p+\epsilon \r) e^{-\Phi} V-\frac{We^{\Phi-\lambda}}{r}\frac{dp}{dr}-\frac{1}{2} \l(p+\epsilon \r) e^{\Phi}H\,, \nonumber \\
    \text{}  \label{eqn:x}
\end{eqnarray}
$m=m(r)$ is the enclosed mass of the star and $\gamma$ is the adiabatic index defined as
\begin{equation}
    \gamma=\frac{\l(p+\epsilon \r)}{p}\l(\frac{\partial p}{\partial \epsilon }\r)\bigg|_{ad} ~.
     \label{eqn:gamma}
\end{equation}

While solving the differential equations Eqs. \eqref{eqn:dh1}-\eqref{eqn:dx} along with the algebraic Eqs. \eqref{eqn:h}-\eqref{eqn:v}, we have to impose proper boundary conditions, i.e., the perturbation functions are finite throughout the interior of the star (particularly at the centre, i.e., at $r=0$) and the perturbed pressure ($\Delta p$) vanishes at the surface.  Function values at the centre of the star can be found using the Taylor series expansion method described in Appendix B of~\cite{Detweiler83} (see also  Appendix A of~\cite{Sotani2001}). The vanishing perturbed pressure at the stellar surface is equivalent to the condition $X(R)=0$ (as, $\Delta p=-r^l e^{-\Phi}X$). We followed the procedure described in ~\cite{Detweiler83} to find the unique solution for a given value of $l$ and $\omega$ satisfying all the boundary conditions inside the star.
\subsubsection{Perturbations outside the star and complex eigenfrequencies}
The perturbations outside the star are described by the Zerilli  equation~\cite{Zerilli}.
\begin{equation}
    \frac{d^2Z}{dr_*^2}+\omega^2 Z=V_Z Z
    \label{eqn:zerilli}
\end{equation}
 where $r_*=r+2m \log \l({\frac{r}{2m}-1}\r)$ is the tortoise co-ordinate and $V_Z$ is defined as ~\cite{Zerilli},
 \begin{eqnarray}
     V_Z&=&\frac{2\l(r-2m\r)}{r^4 \l(nr+3m\r)^2}\Big[n^2(n+1)r^3 \nonumber \\
     &+&3n^2mr^3+9nm^2r+9m^3\Big] ~,
 \end{eqnarray}
where $n=\frac{1}{2} (l+2)(l-1)$. Asymptotically the wave solution to \eqref{eqn:zerilli} can be expressed as \eqref{eqn:zerillisolution},
\begin{eqnarray}
    Z=A(\omega)Z_{in}&+&B(\omega) Z_{out}\,, \label{eqn:zerillisolution}\\
    Z_{out}=e^{-i\omega r^*} \sum_{j=0}^{j=\infty}\alpha_j r^{-j}&,& Z_{in}=e^{i\omega r^*} \sum_{j=0}^{j=\infty}\bar{\alpha}_j r^{-j} ~.\nonumber
    \end{eqnarray}
 Keeping  terms up to $j=2$ one finds,
 \begin{eqnarray}
     \alpha_1&=&-\frac{i}{\omega}(n+1)\alpha_0, \\
     \alpha_2&=&\frac{-1}{2\omega^2}\l[n(n+1)-i3M\omega\l(1+\frac{2}{n}\r)\r]\alpha_0
 \end{eqnarray}

For initial boundary values of Zerilli functions, we use the method described in ~\cite{Fackerell,Detweiler85,Sotani2001}. Setting $m=M$ and perturbed fluid variables to 0 (i.e., $W=V=0$) outside the star, connection between the metric functions \eqref{eqn:metricfunctions} with Zerilli function ($Z$ in Eq.\eqref{eqn:zerilli}) can be written as,
\begin{eqnarray}
    \begin{pmatrix}
    r^lK\\
    r^{l+1}H_1
    \end{pmatrix}
    &=&Q\begin{pmatrix}
    Z \\
    \frac{dZ}{dr_*}
    \end{pmatrix}
    \label{eqn:zerillconnection}
\end{eqnarray}
\begin{eqnarray*}
    Q&=&\begin{pmatrix}
    \frac{ n(n+1)r^2+3nMr+6M^2}{r^2(nr+3M)} & 1\\
    \frac{nr^2-3nMr-3M^2}{(r-2M)(nr+3M)} & \frac{r^2}{r-2M}
    \end{pmatrix} \\
\end{eqnarray*}

The initial boundary values of Zerilli functions are fixed using \eqref{eqn:zerillconnection}. Then, the Zerilli equation \eqref{eqn:zerilli} is integrated numerically to infinity, and the complex coefficients $A(\omega) ,\ B(\omega)$ are obtained matching the analytic expressions for $Z$ and $\frac{dZ}{dr_*}$ with the numerically obtained value of $Z$ and $\frac{dZ}{dr_*}$. The natural frequencies of an oscillating neutron star, which are not driven by incoming gravitational radiation, represent the quasi-normal mode frequencies. Mathematically we have to find the complex roots of  $A(\omega)=0$, representing the complex eigenfrequencies of  QNMs.

\section{Relations for $\Delta f$ and $\Delta \tau$}\label{sec:appendix_delta_f_tau}

In Sec.~\ref{sec:vary_G}, we found fit-relations for $\Delta f$ and $\Delta \tau$ as a function of mass of NS admixed DM and percentage of DM (see Eqns.~\ref{eqn:deltaf_fit_quadratic},~\ref{eqn:deltatau_fit}). We found that $\Delta f \propto \sqrt{M}$ and $\Delta \tau \propto M^{-2}$ with a quadratic and linear dependence on $f_{DM}[\%]$, respectively. We reported the fit coefficients in Table~\ref{table:deltaf_deltatau_fits}. However, here the hadronic EoS was fixed. We need to verify i) whether the relations (Eqns.~\ref{eqn:deltaf_fit_quadratic},~\ref{eqn:deltatau_fit}) hold when the hadronic EoS is changed and if it does, ii) whether the fitting coefficients ($C_{fi}$ and $C_{\tau}$) change.

We had used `Hadronic' parametrization before (see Table.~\ref{table:parameters}) with $m^*/m=0.68$. To change the hadronic EoS, we choose two additional values of $m^*/m$ (0.63 and 0.65) and check the relations since $m^*/m$ is known to control the stiffness of the EoS and is the most dominant nuclear empirical parameter. This is the only reason why we consider different values of $m^*/m$ adn these is nothing special about the chosen values. We only consider quadratic fit for $\Delta f$ as it was seen to be a better fit, having an accuracy under $5\%$. 

\begin{figure}
    \centering
    \includegraphics[width=0.95\linewidth,trim={0 0.5cm 0 0}, height=0.9\linewidth]{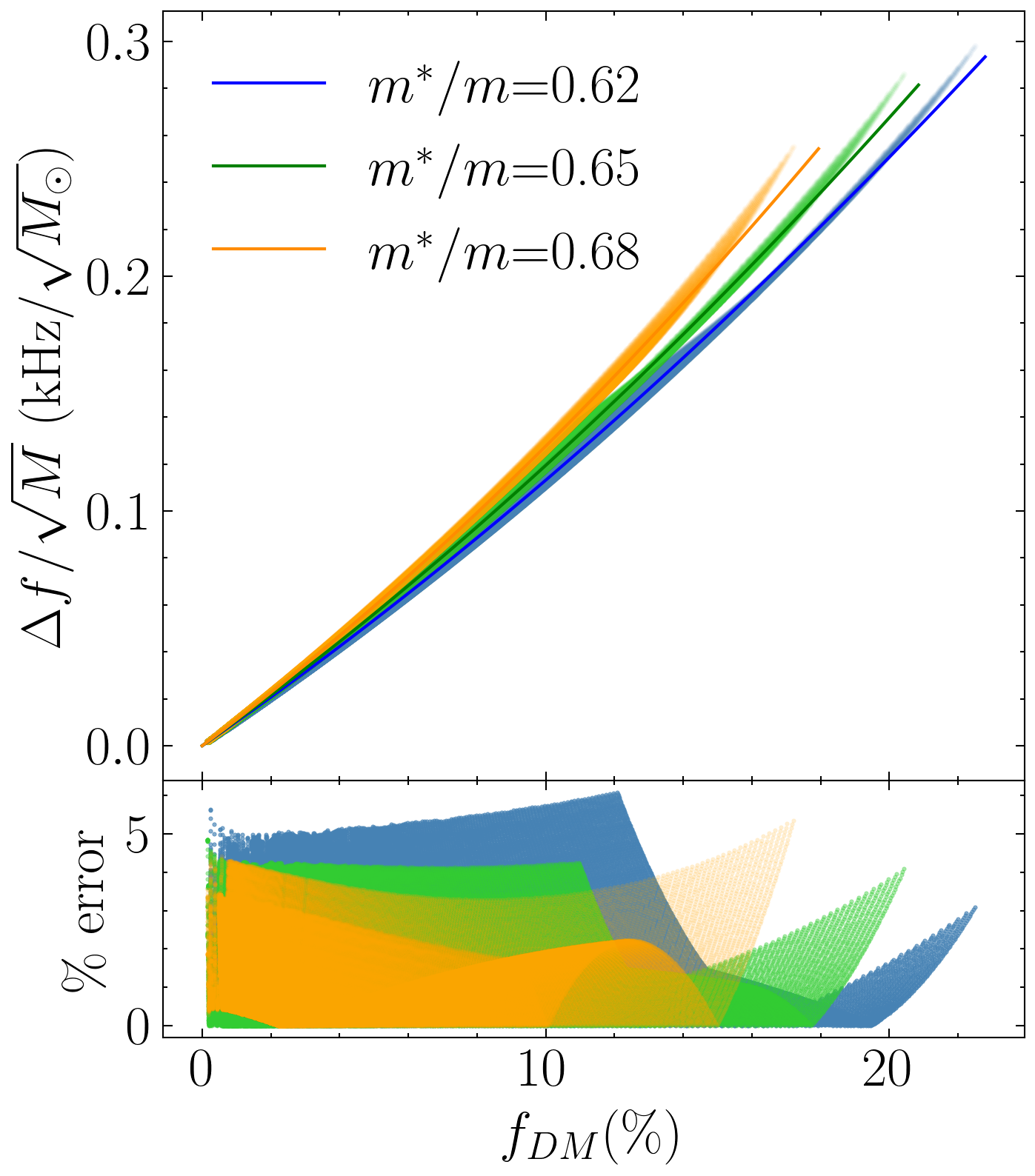}
    \caption{The top panel shows $\Delta f(M)/\sqrt{M}$ as a function of $f_{DM}$ obtained by varying $G$. Three scatter plots correspond to three values of $m^*/m$. The other nuclear parameters are fixed to `Hadronic' (refer Table.~\ref{table:parameters}). Curves of of the same color are the best-fit curves to the corresponding scatter plot given by Eq.~(\ref{eqn:deltaf_fit_quadratic}). The fit coefficients are reported in Table~\ref{table:deltaf_deltatau_vary_mstar}. The bottom panel shows percent error for the fits.}
    \label{fig:deltaf_vs_fdm_vary_mstar}
\end{figure}

\begin{figure}
    \centering
    \includegraphics[width=0.95\linewidth,trim={0 0.5cm 0 0}, height=0.9\linewidth]{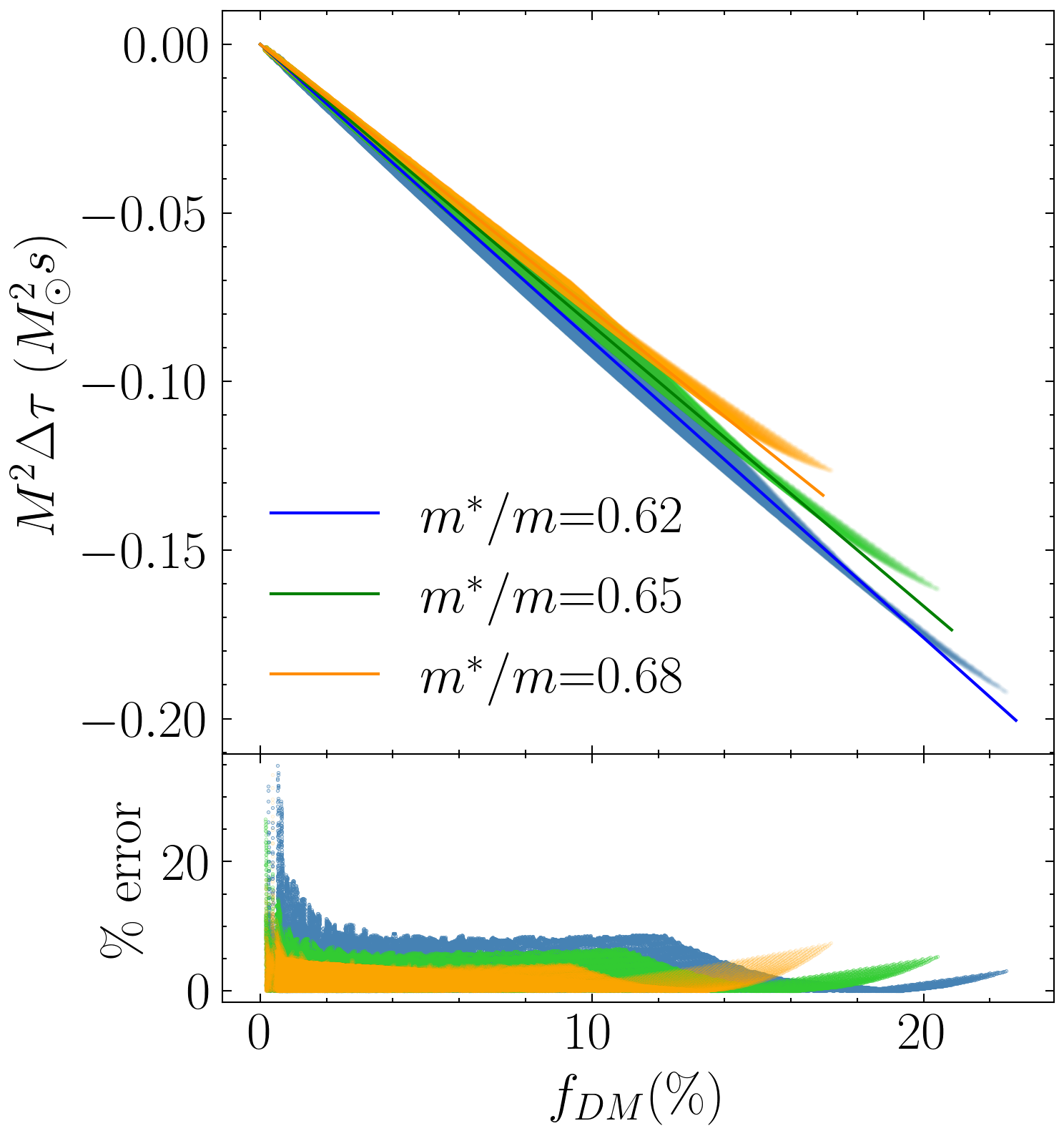}
    \caption{The top panel shows $M^2\Delta \tau(M)$ as a function of $f_{DM}$ obtained by varying $G$. Three scatter plots correspond to three values of $m^*/m$. The other nuclear parameters are fixed to `Hadronic' (refer Table.~\ref{table:parameters}). Curves of of the same color are the best-fit curves to the corresponding scatter plot given by Eq.~(\ref{eqn:deltatau_fit}). The fit coefficients are reported in Table.~\ref{table:deltaf_deltatau_vary_mstar}. The bottom panel shows percent error for the fits.}
    \label{fig:deltatau_vs_fdm_vary_mstar}
\end{figure}

We plot $\Delta f/M$ and $M^2\Delta \tau$ as a function of $f_{DM}[\%]$ in Fig,~\ref{fig:deltaf_vs_fdm_vary_mstar} and Fig.~\ref{fig:deltatau_vs_fdm_vary_mstar} respectively. Blue color represents $m^*/m=0.62$, green color represents $m^*/m=0.65$, and orange color represents $m^*/m=0.68$, which is the same case as discussed in Sec.~\ref{sec:vary_G}. Lower $m^*/m$ values lead to stiffer EoS, allowing for larger DM fractions, as can be seen in the figure. We find that the relations Eqs.~(\ref{eqn:deltaf_fit_quadratic}) and (\ref{eqn:deltatau_fit}) hold for these as well. All these fits agree within $\sim 5\%$ accuracy. The fit coefficients, however, change and are tabulated in Table~\ref{table:deltaf_deltatau_vary_mstar}. For the case $\Delta f$, the slope is higher for softer EoS, i.e. the change in $f$-mode frequency $f$ for a fixed value $f_{DM}$ is higher in the case of softer EoS. The opposite effect is seen in case of damping time $\tau$. The decrease in $f$-mode damping time is less in the case of soft EoS for a given fraction of DM in NS.

We conclude that the relation of Eqs.~(\ref{eqn:deltaf_fit_quadratic}) and (\ref{eqn:deltatau_fit}) hold for different hadronic EoSs. We also conclude that i) $\Delta f$ grows as $\sqrt{M}$ and quadratically with $f_{DM}$ as $(af_{DM}+bf_{DM}^2)$, and ii) $\Delta \tau$ is proportional to $M^{-2}$ and decreases linearly with $f_{DM}$. The fitting coefficients depend on the hadronic EoS; hence, the fit is not a universal relation.

\begin{table}
    \caption{\label{table:deltaf_deltatau_vary_mstar}%
    Fitting coefficients for Eqs.~(\ref{eqn:deltaf_fit_quadratic}) and (\ref{eqn:deltatau_fit}). $C_{fi}$ are given in units of kHz/$\sqrt{M_{\odot}}$, $C_{\tau}$ in units of $M_{\odot}^2$s. $m^*/m$ is varied and the rest of the nuclear parameters used are from the setup `Hadronic' (see Table~\ref{table:parameters}).}
    \begin{ruledtabular}
    \begin{tabular}{c|cc|c}
       $m^*/m$& $C_{f2} [\times 10^{-2}]$ & $C_{f3} [\times 10^{-4}]$ & $C_{\tau} [\times 10^{-3}]$\\ \hline
        0.62 & 1.01 $\pm$ 0.11& 1.22 $\pm$ 0.87 & -8.80 $\pm$ 0.39\\
        0.65 & 1.05 $\pm$ 0.12& 1.44 $\pm$ 1.03 & -8.33 $\pm$ 0.42\\
        0.68 & 1.10 $\pm$ 0.14& 1.78 $\pm$ 1.42 & -7.88 $\pm$ 0.49\\
    \end{tabular} 
    \end{ruledtabular}
\end{table}

\section{Posterior Distributions}\label{sec:appendix_posterior}

\begin{figure*}
    \includegraphics[width=\linewidth,trim={0 0.8cm 0 0}]{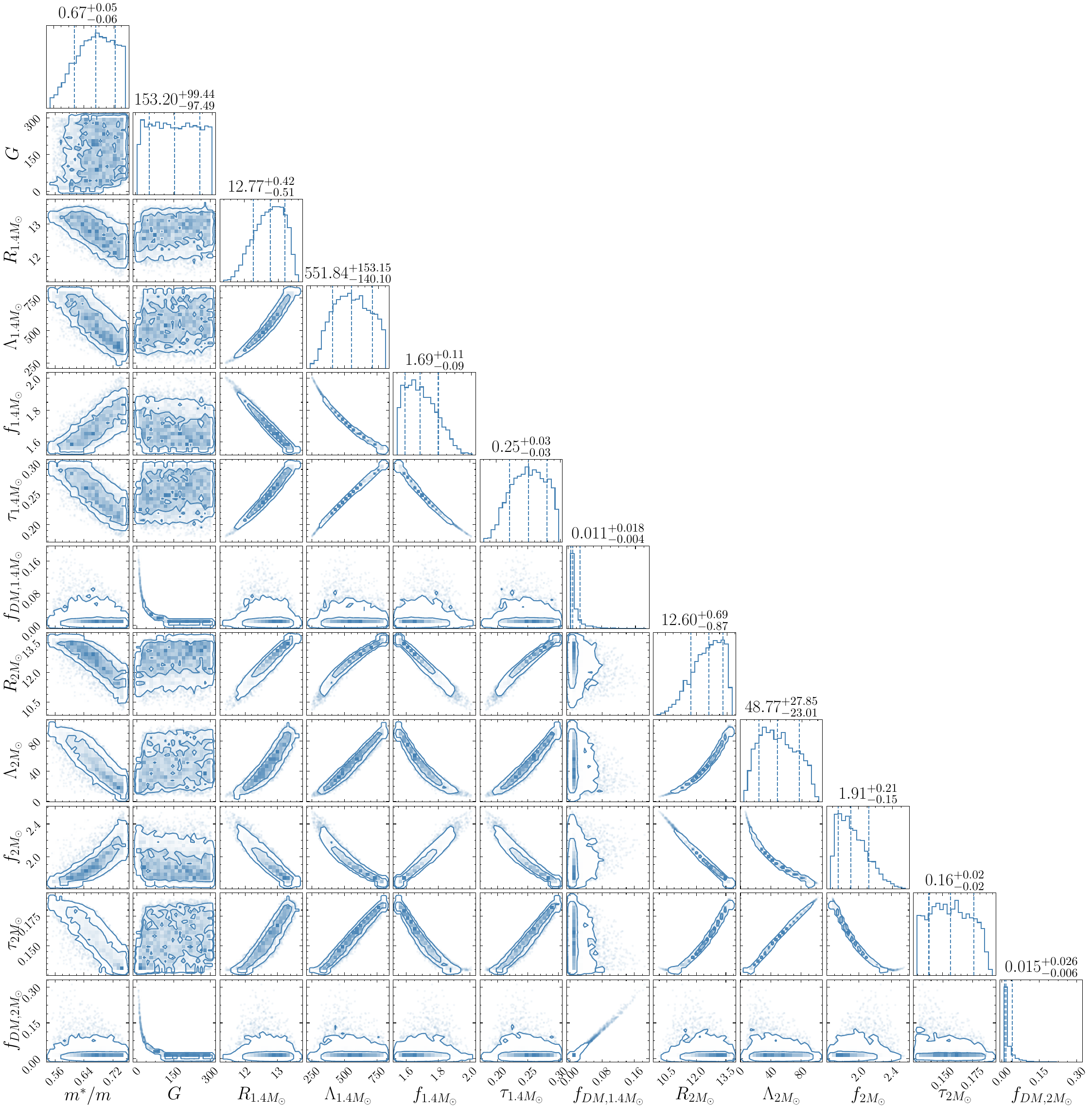}
    \caption{Corner plot showing posteriors of select parameters namely: effective mass ($m^*/m)$, DM self-interaction ($G$), NS observables ($R_{{1.4M_{\odot}}}$, $\Lambda_{{1.4M_{\odot}}}$, $R_{{2.0M_{\odot}}}$, $\Lambda_{{2.0M_{\odot}}}$), DM fraction ($f_{DM, 1.4M_{\odot}}$, $f_{DM, 2,M_{\odot}}$), and $f$-mode characteristics ($f_{{1.4M_{\odot}}}$, $\tau_{{1.4M_{\odot}}}$, $f_{{2M_{\odot}}}$, $\tau_{{2M_{\odot}}}$). Posteriors are obtained  after applying the $\chi EFT$, GW170817 and $2M_{\odot}$ constraints. The vertical lines and values denote the median and the middle $68\%$ range of the distribution. The parameter range is given in~\cref{table:parameters}}
    \label{fig:corner_allvary}
\end{figure*}

In order to understand the correlations better, we plot the posterior distribution of the effective mass ($m^*/m$) DM self-interaction parameter ($G$), NS observables ($R_{{1.4M_{\odot}}}$, $\Lambda_{{1.4M_{\odot}}}$, $R_{{2.0M_{\odot}}}$, $\Lambda_{{2.0M_{\odot}}}$), DM fraction ($f_{DM, 1.4M_{\odot}}$, $f_{DM, 2, M_{\odot}}$), and $f$-mode characteristics ($f_{{1.4M_{\odot}}}$, $\tau_{{1.4M_{\odot}}}$, $f_{{2M_{\odot}}}$, $\tau_{{2M_{\odot}}}$) obtained after applying all the filters ($\chi EFT$, $2M_{\odot}$ and GW170817) in Fig.~\ref{fig:corner_allvary}. The vertical lines in the 1D distribution denote the middle $68\%$ range. We make the following observation:

\begin{itemize}
    \item $m^*/m$ shows correlation with all the parameters shown except with DM parameters: $G$, and DM fractions ($f_{DM, 1.4M_{\odot}}$ and $f_{DM, 2, M_{\odot}}$).
    \item $G$ is not seen to be constrained after applying all the filters and remains uncorrelated except in the case with DM fraction. We observe an inverse relation of $f_{DM}$ with $G$, which is explored in more detail in Sec.~\ref{sec:vary_all}.
    \item $R$, $\Lambda$, $f$, and $\tau$ exhibit tight relations among themselves. $\Lambda$ is known to depend of $R$ though the equation $\Lambda = \frac{2}{3}\frac{k_2}{C^5}$. We showed that the mass-scaled $f$-mode characteristics follow a tight relation with $\Lambda$ as given by Eqn.~\ref{eqn:UR_lam}. Combining these relations, we expect the $f$-mode characteristics to be related to the radius.
    \item The DM fraction of both 1.4$M_{\odot}$ and 2$M_{\odot}$ DM admixed NS are restricted to lower values resulting in positively skewed distribution peaking at $f_{DM}=0$. The 90$\%$ quantile for $f_{DM, 1.4M_{\odot}}$ and $f_{DM, 2M_{\odot}}$ is $3.97\%$ and $5.79\%$, respectively. This suggests that the current constraints favor a lower DM fraction. A tight relation is seen between $f_{DM, 1.4M_{\odot}}$ and $f_{DM, 2M_{\odot}}$. This is in accordance with the relation~\ref{eqn:dmfrac_fit_linear} explored in detail earlier in this work.
\end{itemize}

We also check the effect of imposing a larger maximum mass constraint. A recent analysis of the black widow pulsar, PSR J0952-0607~\cite{Romani2022}, resulted in a high pulsar mass of $M=2.35 \pm 0.17 M_{\odot}$. However, this system is very rapidly rotating with a period of $P = 1.41$~ms, which means that the lower limit imposed by this on the maximum mass of non-rotating stars would be lower than $2.35 M_{\odot}$~\cite{BreuRezzolla2016}. To check the effect of a higher $M_{max}$ constraint, we checked the effect on posteriors if the maximum mass limit of NS were $2.3 M_{\odot}$. The plot is not shown here. We see the following differences:
\begin{itemize}
    \item The posteriors show the same qualitative features, only the ranges change.
    \item Higher values of $m^*/m$ and lower values of $G$ are unfavoured. This is expected as these result in softening of EoS and a higher mass limit filters these out.
    \item $G$ remains unconstrained. 
    \item Low values of $R$, $\Lambda$, $\tau$ and high values of $f$ get filtered out as expected.
    \item The DM fraction remains positively skewed with peak at $0\%$. The 90$\%$ quantile for $f_{DM, 1.4M_{\odot}}$ and $f_{DM, 2M_{\odot}}$ reduces to  $3.03\%$ and $4.29\%$, respectively. This is because now, that a higher maximum mass is required, higher DM fractions are unfavoured as they soften the EoS. So observation of heavier NSs is a way to rule out the presence of DM.
\end{itemize}



\section{Fixed $m^*/m$}\label{sec:appendix_fixed_mstar}

\begin{figure}
\includegraphics[width=\linewidth,trim={0 0.8cm 0 0}]{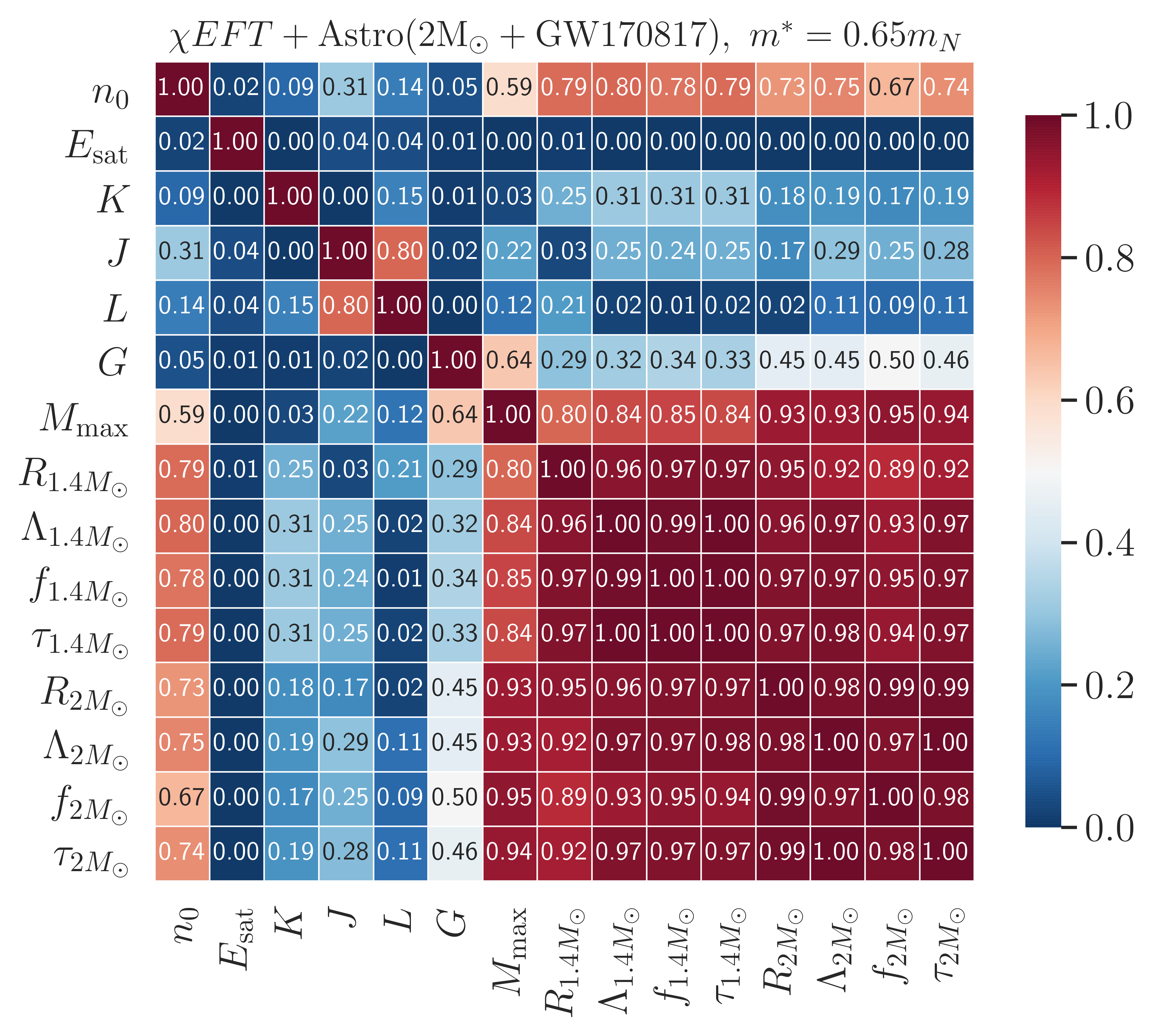}
    \caption{ Correlation matrix showing the correlations among the nuclear parameters, DM interaction parameter, NS observables and the f-mode characteristics. Correlations are obtained  After applying the $\chi EFT$, GW170817 and $2M_{\odot}$ constraints. $m^*/m$ is fixed to 0.65. The range of rest of the parameters is given in~\cref{table:parameters}}
    \label{fig:correlation_fix_mstar_065}
\end{figure}

\begin{figure}
\includegraphics[width=\linewidth,trim={0 0.8cm 0 0}]{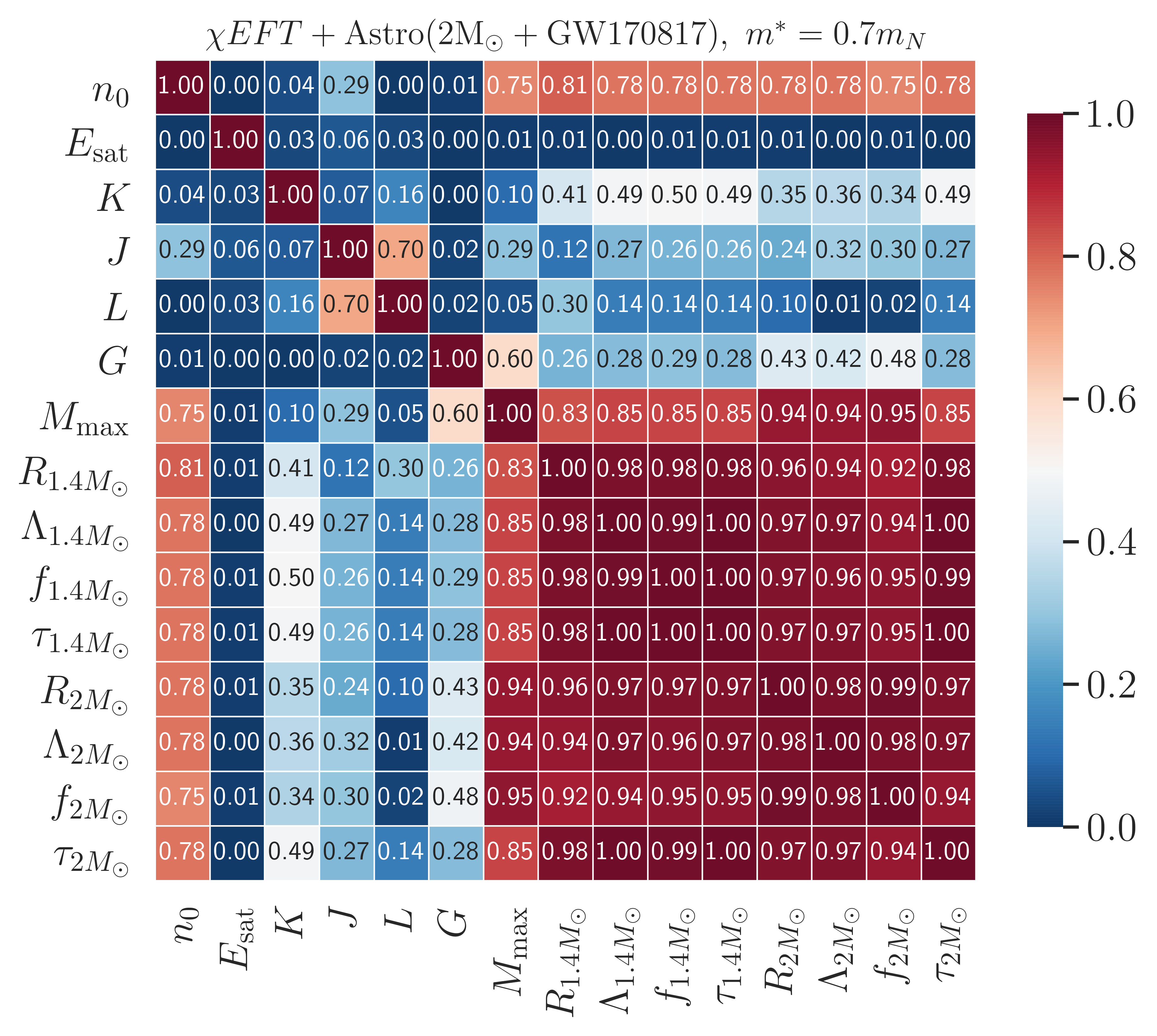}
    \caption{ Correlation matrix showing the correlations among the nuclear parameters, DM interaction parameter, NS observables and the f-mode characteristics. Correlations are obtained  After applying the $\chi EFT$, GW170817 and $2M_{\odot}$ constraints. $m^*/m$ is fixed to 0.7. The range of rest of the parameters is given in~\cref{table:parameters}}
    \label{fig:correlation_fix_mstar_070}
\end{figure}

We check the correlations here, keeping the effective mass fixed to three values: 0.6, 0.65, and 0.7. We plot the correlations for these cases in in Fig.~\ref{fig:correlation_fix_mstar_060}, Fig.~\ref{fig:correlation_fix_mstar_065} and Fig.~\ref{fig:correlation_fix_mstar_070} respectively. We draw the following conclusions from these plots:

\begin{figure}
    \includegraphics[width=\linewidth,trim={0 0.8cm 0 0}]{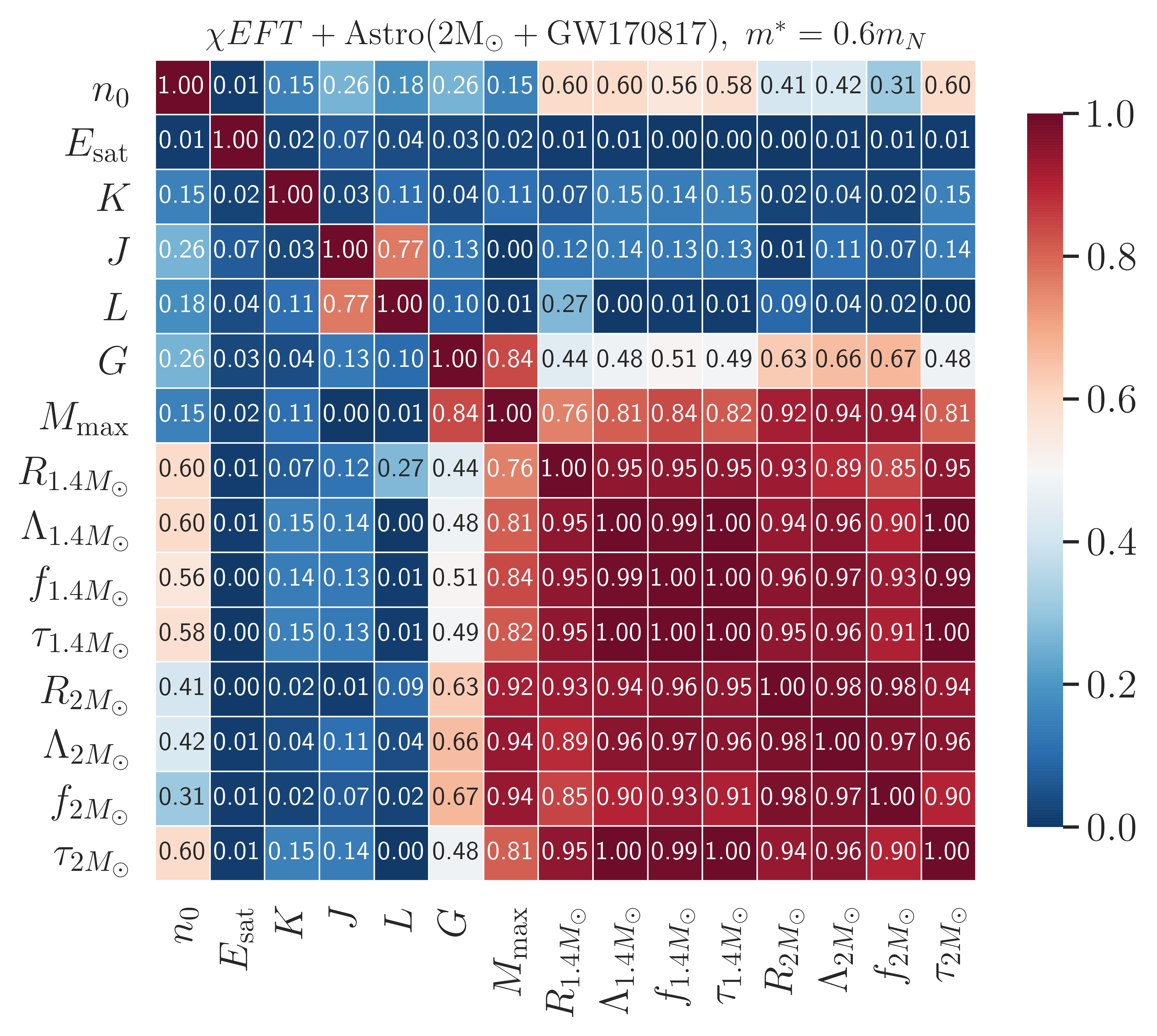}
    \caption{ Correlation matrix showing the correlations among the nuclear parameters, DM interaction parameter, NS observables and the f-mode characteristics. Correlations are obtained  After applying the $\chi EFT$, GW170817 and $2M_{\odot}$ constraints. $m^*/m$ is fixed to 0.6. The range of rest of the parameters is given in~\cref{table:parameters}}
    \label{fig:correlation_fix_mstar_060}
\end{figure}

\begin{itemize}
    \item We find an emergence of correlations of NS observables with $n_0$ and $G$. These are the next dominant parameters after the effective mass. The correlation of $G$ with $M_{max}$ is higher than the other observables.
    \item For $m^*/m=0.6$, $G$ is moderately correlated with all the NS observables and strongest with $M_{max}$ (0.84). The maximum mass is dictated by $G$ alone. The correlation with 2$M_{\odot}$ properties is larger than that of 1.4$M_{\odot}$. This shows that $G$ has a greater effect at high densities. All other nuclear parameters are uncorrelated.
    \item Correlation of $G$ reduces with increasing $m^*/m$ and that of $n_0$ increases. This is because lower effective mass leads to stiffer EoS, and $G$ is known to soften it. Since we add a cut of $2M_{\odot}$, the already soft EoS (higher $m^*/m$) gets filtered out upon adding DM. Hence, we get a higher correlation for lower effective mass.
    \item Nuclear parameters show moderate correlation with NS observables as we increase $m^*/m$. $E_{sat}$ stays completely uncorrelated ($\approx 0$) in all the cases.
    \item All NS observables remain strongly correlated with each other.
\end{itemize}

The effect of $G$ is only to soften the EoS. Hence, for larger values of effective mass, when the maximum mass of the purely hadronic NS is already low, $G$ cannot have much impact since we add a $2M_{\odot}$ cut-off. This explains the reduction in correlations of $G$ as $m^*/m$ is increased.